\documentclass[journal]{IEEEtran}
\usepackage{amsmath,amssymb,amsfonts,mathrsfs,mathtools,bm, bbm, dsfont,mathrsfs,amsthm,blkarray}
\usepackage[dvipdfm]{graphicx}
\usepackage[dvipsnames]{xcolor}

\usepackage{epsfig,epsf,psfrag,latexsym}
\usepackage{graphicx}
\usepackage[ruled,vlined]{algorithm2e}
\include{pythonlisting}

\usepackage{booktabs}
\usepackage{multirow,multicol}

\usepackage[breaklinks,colorlinks, linkcolor=MidnightBlue, anchorcolor=MidnightBlue, citecolor=MidnightBlue, urlcolor=MidnightBlue]{hyperref}
\usepackage[hyphenbreaks]{breakurl}
\usepackage{cite}

 \def\old#1{}    % Please don't remove this... This command includes the text to be deleted.

\def\nn{\nonumber}
\def\beq{\begin{equation}}
\def\eeq{\end{equation}}
\def\bea{\begin{eqnarray}}
\def\eea{\end{eqnarray}}
\def\ba{\begin{array}}
\def\ea{\end{array}}

\def\bitem{\begin{itemize}}
\def\eitem{\end{itemize}}
\def\ben{\begin{enumerate}}
\def\een{\end{enumerate}}

\def\ie{{\it i.e.,\ \/}}

\newcommand{\Amsc}{\mathscr{A}}

\def\dbf{{\bm d}}
\def\ebf{{\bm e}}

\def\gbf{{\bm g}}
\def\hbf{{\bm h}}

\def\lbf{{\bm l}}

\def\pbf{{\bm p}}
\def\qbf{{\bm q}}

\def\vbf{{\bm v}}

\def\Abf{{\bm A}}

\def\Dbf{{\bm D}}

\def\Lbf{{\bm L}}

\def\Hc{{\cal H}}

\def\Pc{{\cal P}}

\def\cross{\!  \times  \!}

\newcommand{\beqa}{\begin{eqnarray}}
\newcommand{\eeqa}{\end{eqnarray}}
\newcommand{\beqan}{\begin{eqnarray*}}
\newcommand{\eeqan}{\end{eqnarray*}}

\newcommand{\Rset}{\mathbb{R}}
\renewcommand{\v}[1]{{\bm{#1}}}

\newcommand{\ol}[1]{\ensuremath{\overline{{#1}}}}
\newcommand{\ul}[1]{\ensuremath{\underline{{#1}}}}

\newcounter{l1}
\newcounter{l2}
\newcounter{l3}
\newcommand{\bdotlist}{\begin{list}{$\bullet$}{}}
\newcommand{\bboxlist}{\begin{list}{$\Box$}{}}
\newcommand{\bbboxlist}{\begin{list}{\raisebox{.005in}{{\tiny
$\blacksquare$ \ \ }}}{}}
\newcommand{\bdashlist}{\begin{list}{$-$}{} }
\newcommand{\blist}{\begin{list}{}{} }
\newcommand{\barablist}{\begin{list}{\arabic{l1}}{\usecounter{l1}}}
\newcommand{\balphlist}{\begin{list}{(\alph{l2})}{\usecounter{l2}}}
\newcommand{\bAlphlist}{\begin{list}{\Alph{l2}.}{\usecounter{l2}}}
\newcommand{\bdiamlist}{\begin{list}{$\diamond$}{}}
\newcommand{\bromalist}{\begin{list}{(\roman{l3})}{\usecounter{l3}}}

\title{Enhancing Microgrid Resilience with Green Hydrogen Storage}

\author{Shreshtha Dhankar~\IEEEmembership{Student Member,~IEEE,}
\quad
Cong Chen ~\IEEEmembership{Student Member,~IEEE,}
\quad Lang~Tong~\IEEEmembership{Fellow,~IEEE}
\thanks{\scriptsize
Shreshtha Dhankar (sd728@cornell.edu) is with the School of Chemical Engineering, Cornell University, Ithaca NY, USA. Cong Chen and Lang Tong (\{cc2662, lt35\}@cornell.edu) are with the School of Electrical and Computer Engineering, Cornell University, Ithaca NY, USA. (Corresponding author: Cong Chen)}
\thanks{\scriptsize  The work is supported by the National Science Foundation under
Award  2218110, and Power Systems and Engineering Research Center (PSERC) Research Project M-46.}
}

\begin{document}
% \vspace{-cm}
\maketitle
\begin{abstract}

% We consider the problem of hydrogen storage integration in microgrids to improve the electricity supply resilience. Nonlinear electrochemical models of electrolyzers and fuel cells for hydrogen storage are considered through a piecewise linear approximation. This maintains an accurate and computationally efficient  hydrogen storage model when operating the microgrid in the real-time model predictive control. The piecewise linear approximation method has simulation results approaching the ideal benchmark which is the nonlinear hydrogen model with perfect forecasts. We also investigate the influence on different microgrid resilience metrics caused by various penalty functions for the loss-of-load. We show that an L1-norm penalty works well for minimizing total loss load, L2-norm penalty reduces the time duration-of-outage, and the weighted sum of infinite-norm and L1-norm penalties reduces the maximum loss-of-load.

We consider the problem of hydrogen storage integration in microgrids to improve the electricity supply resilience. Nonlinear effects from electrochemical models of electrolyzers and fuel cells for hydrogen storage are considered, making scheduling under the nonlinear model intractable and the conventional linear approximation infeasible. A piecewise linear model approximation with feasibility projection is proposed, resulting in a computationally efficient model predictive control for hydrogen storage operation. Several resilience performance measures, such as loss-of-load, duration-of-outage, and system cost, are used in performance evaluation. Simulations for the proposed optimization demonstrated a 13\%-48\% reduction in duration-of-outage, a 6.4\%-21.7\% reduction in system cost, and a 95\% reduction in loss-of-load for critical loads compared to the scheduling algorithm involving linear model approximation. The performance gap of the proposed optimization to the benchmark involving the accurate nonlinear electrochemical model is less than 1\% in most metrics.

\end{abstract}


\section{Introduction}\label{sec:Intro}
% The variability in renewable power production poses a significant challenge to the resilience of microgrids when integrating large-scale rooftop solar. Green hydrogen storage offers a potential solution to enhance system reliability and resilience. However, the full extent of its capabilities in addressing these challenges remains unknown, primarily due to unexplored synergies between solar resource variations, contingency events, and hydrogen production schedules.

% Hydrogen storage emerges as an ideal candidate for long-term energy storage, owing to its high energy density and negligible self-discharge rate when compared to conventional batteries \cite{zhang2017comparative}.  In the microgrid's regular operational state, extra electricity from renewables is harnessed by electrolyzer for `green' hydrogen production. This green hydrogen is stored as a strategic reserve that can be used during contingency events (eg. outages caused by snowstorms) by employing fuel cells to convert the hydrogen back into electricity. This approach enhances the resilience of the microgrid during islanding mode when the external grid cannot provide support. %, thus improving the microgrid's ability to weather such events.

The large-scale integration of intermittent renewable resources and the increasing occurrence of severe climate events pose significant challenges to the resilience of the modern power grid.  It has been widely recognized that microgrids operating at both grid-connected and islanding modes are promising ways to reduce customer outages and enhance overall resilience. To this end, a synergistic operation of local renewable generation, storage charging/discharging, and prioritized scheduling of flexible demand is critical in minimizing the loss-of-load and outage duration.

This work focuses on the role of green hydrogen storage for microgrid resilience. Compared with electric battery systems, hydrogen storage is a strong candidate for long-duration energy storage owing to its high energy density and negligible self-discharge rate \cite{zhang2017comparative}. Surplus renewable can be harnessed by electrolyzers to produce green hydrogen. With advances in electrolyzer and fuel cell technologies and the economy of scale, green hydrogen is poised to be a dual-use technology critical to the economic and resilient operations of the decarbonized power and transportation systems. During events when power from the parent grid is lost or severely curtailed, the microgrid can convert green hydrogen back into electricity with fuel cells to enhance the system resilience. 

% rely on the strategic reserve of green hydrogen and 
% \tcr{(add a few references here if appropriate)}. 
% Green hydrogen can be stored as a strategic reserve. 
% This work focuses on the role of green hydrogen storage for microgrid resilience.  Compared with electric battery systems, hydrogen storage emerges as an ideal candidate for long-term energy storage owing to its high energy density and negligible self-discharge rate \cite{zhang2017comparative}. In the microgrid’s regular operational state, surplus renewables can be harnessed by an electrolyzer and stored as a strategic reserve for severe events, with fuel cells converting the green hydrogen back into electricity. 
\vspace{-0.4cm}
\subsection{Summary of contribution}
  
% We adopt model predictive control (MPC) to co-optimize microgrid and hydrogen storage in real-time to enhance the energy system resilience. Here are our main contributions.
We develop novel modeling and optimization techniques to enhance microgrid resilience, taking into account the nonlinear electrochemical characteristics of electrolyzers and fuel cells.

First, we propose a resilience-enhancing energy management system (EMS) for a microgrid capable of operating at both grid-connected and islanding modes.  The EMS optimizes the scheduling of prioritized demand, distributed energy resources (eg. roof-top solar), and electrolyzer and fuel cell operations under several objective functions involving loss-of-load norms as resilience metrics. 

Second, we propose a piecewise linear approximation of the nonlinear electrochemical hydrogen storage model and a feasibility projection approach, resulting in a significantly reduced computation complexity. This makes it possible for the real-time operations of microgrid and hydrogen storage with a model predictive control (MPC) implementation. 

Third, we present numerical simulation results, demonstrating performance gain over the assumed conventional linear hydrogen charging-discharging model. In particular, we show that the piecewise linear approximation resulted in 
less than 1\% gaps in system cost and total loss-of-load when compared with the accurate but more expensive nonlinear model. Compared with the state-of-the-art linear model, on the other hand, the proposed piecewise linear approximation solution reduces duration-of-outage by 13\%-48\%, system cost by 6.4\%-21.7\%, and the loss-of-load by 95\%. 

Our simulation also provided insights into the effectiveness of different loss-of-load penalty functions used in the objective function.  Specifically, we showed that the $l_1$-norm penalty function worked best in reducing
total loss-of-load. The mixed norm (i.e., a weighted sum of $l_1$ and $l_{\infty}$ norms) was preferred if both the maximum loss-of-load and the total loss-of-load were important resilience metrics. And $l_2$-norm
penalty outperformed others in shortening the duration-of-outage.

% Second, we explore microgrid optimization toward different resilience metrics via various penalty functions for the loss-of-load. The $l_{1}$-norm penalty function works best in reducing total loss-of-load. The mixed norm (\ie a weighted sum of  $l_{1}$ and $l_{\infty}$ norms) is preferred if both maximum loss-of-load and the total loss-of-load are important resilience metrics. $l_{2}$-norm penalty outperforms in shortening the duration-of-outage.% and maintaining balanced resilience metrics for customers categorized by different values of load shedding penalty.

% Third, the proposed optimization closely approximates the simulation results of the ideal benchmark nonlinear model with differences less than 1\% in system cost and total loss-of-load. Compared to the linear approximation, the proposed method reduces duration-of-outage by 13\%-48\%, system cost by 6.4\%-21.7\%, and the loss-of-load for critical loads by 95\%.
\vspace{-0.4cm}
\subsection{Related work}

% \tcr{Cong: 1) could you check if my writing is correct and if I’m citing the correct paper? 2) What is the approximation in type II like [11], is it still using constant? If not, how is it different from our piecewise linear approximation? explain what kind of problem does this Type II model used. Shreshtha: not sure what you mean by using constant, [11] summarizes the different resilience metrics. 3) I’ll need your help to add the references related since you’re more familiar than me in reading those related papers. We only cite good and relevant paper that we read and understand. 3) are there any paper using two-norm or infinite-norm in the objective function Shreshtha: Not nay to my knowledge. 4) are there any paper considering infeasibility correction method for the approximation model of hydrogen storage? Shreshtha: For this, there aren't any papers. 5) could you check some literature doing rotate outage? And check how they shed load among different categories of customers?}

The hydrogen storage model is critical to this resilience-enhancing EMS. We summarize three types of hydrogen storage models. Type I establishes a linear hydrogen storage model with constant charging/discharging efficiency \cite{tobajas22AE,haggi2022proactive}. Such a linear model relies on the manufacturer's efficiency specifications or linear regression for power-hydrogen relationships. The model simplicity enables efficient real-time decisions in bulk system operation and control problems \cite{tobajas22AE}. However, this approach tends to overlook the crucial electrochemical dynamics of hydrogen storage.  Type II approximates the nonlinear electrochemical models of hydrogen storage with piecewise linear models \cite{Gabrielli16EEEIC, FLAMM21AE}. Such a method balances the model's precision and computation burden.  %\tcr{Such a model accelerates the computation efficiency for system-level simulation and improves the modeling accuracy of hydrogen storage.} 
Type III directly employs nonlinear electrochemical models \cite{bessarabov16pem, larminie03fuel, abomazid21TII,correa2004electrochemical} to conduct simulations on device-level research for hydrogen storage. Such a model is too complicated to be included in a system-level optimization. %, although it considers the nonlinear hydrogen models. 

Our method belongs to Type II. We conduct piecewise linearization for power-hydrogen relationships of the hydrogen storage.  Existing models in  Type II usually characterize hydrogen by volumetric flow rate ($m^3/s$) \cite{Gabrielli16EEEIC}. But the dependence of gas volume on operating conditions (eg. temperature, and pressure)\footnote{Described by the \href{https://en.wikipedia.org/wiki/Ideal_gas_law}{Ideal Gas Law}.} poses challenges in monitoring hydrogen gas volume across the entire operation cycle including electrolyzers, storage tanks, and fuel cells. To address this, our approach considers {\em mass flow rate} ($kg/s$) as it offers a consistent measure for hydrogen, unaffected by variations in the operating conditions. 
%In \cite{FLAMM21AE}, the piecewise linear model is computed with an experimental data driven method under certain operation condition. 
Additionally, most hydrogen storage approximation methods \cite{haggi2022proactive,tobajas22AE, Gabrielli16EEEIC, FLAMM21AE} ignore the issue that they may produce infeasible dispatch signal resulted from  approximation errors. We resolve this issue by proposing a real-time feasibility projection embedded in MPC.

% Our method belongs to Type II. We conduct piecewise linearization for power-hydrogen relationships of both electrolyzer and fuel cells.  In \cite{Gabrielli16EEEIC}, hydrogen storage is originally characterized by hydrogen's volumetric flow rate ($m^3/s$). However, due to the dependence of gas volume on temperature, and pressure\footnote{Described by the \href{https://en.wikipedia.org/wiki/Ideal_gas_law}{Ideal Gas Law}.}, accurately monitoring hydrogen gas volume across the entire operation cycle which includes production using electrolyzers, storage in tanks, and generation using fuel cells poses challenges. To address this, our approach considers {\em mass flow rate} ($kg/s$) as it offers a consistent measure, with mass conservation being unaffected by variations in the operating conditions of devices such as electrolyzers and fuel cells. 
%In \cite{FLAMM21AE}, the piecewise linear model is computed with an experimental data driven method under certain operation condition. 
% Additionally, most hydrogen storage approximation methods \cite{eriksson2017optimization,pan20TSE,tobajas22AE, Gabrielli16EEEIC, FLAMM21AE} ignore the infeasible dispatch signal caused by model approximation errors. We propose a real time infeasibility correction embedded in MPC.

%Numerous approaches have emerged for quantifying resilience, yet, currently, there is no standard resilience metric\cite{panteli2017metrics}. 
When conducting optimization for resilience, most research considers minimizing the total loss-of-load \cite{yao2022quantitative,NAN201735}, and other resilience metrics like minimum load the system can sustain\cite{NAN201735,panteli2017metrics}, duration-of-outage \cite{panteli2017metrics}, percentage of customers experiencing an outage\cite{NAN201735} are sometimes adopted for system evaluations.  In \cite{hussain2017resilience}, different values of lost load were used to indicate the priority of load shedding. Here, we consider customers with different values of lost load and penalty function designs toward different resilience metrics for the microgrid scheduling during the contingency period.% \tcr{In this work, we only focus on the outage phase and not the preparation phase.}%, including the $l_{2}$-norm and the mixed-norm, which is a weighted sum of $l_{1}$-norm and infinite-norm. 

\vspace{-0.2cm}
\section{Electrochemical model of hydrogen storage}
\label{sec:A_Access}
In this section, we first describe nonlinear models for electrolyzers and fuel cells\cite{abomazid21TII,correa2004electrochemical}, and then introduce the piecewise linear approximation model for hydrogen storage.
% \subsection{Faraday's Law of Electrolysis}
% Faraday's Law of Electrolysis gives the relationship between mass and the amount of electric charge. Consider an arbitrary half-cell reaction: 

%  $M \rightarrow M^{++} + ne^{-}$
\vspace{-0.3cm}
\subsection{Electrochemical model of electorlyzer}

An electrolyzer uses electricity (current) to split water into hydrogen and oxygen. The most widely used type of electrolyzer is the Proton Exchange Membrane (PEM) electrolyzer owing to its fast response time {\cite[p.5]{bessarabov16pem}. We adopt the model for the PEM electrolyzer cells developed by Abomazid et al. in \cite{abomazid21TII}, which takes into account the effect of operating temperature, pressure, and the number of identical cells.  %It should be noted that while these analytical relationships were developed for a single PEM cell the proposed approach could be extended to electrolyzer stacks consisting of a number of identical cells in series. 

Let $t$ be the index for time intervals. We denote the electrolyzer current as $ I^e_t (A)$ and the output hydrogen flow rate as $h^{e}_t (kg/s)$.\footnote{Superscript `$e$' represents electrolyzer rather than power of a number.} From Faraday's Law of Electrolysis {\cite[p.20]{bessarabov16pem}}, the current input and hydrogen flow output relationship of the electrolyzer is linear given by 
% The relationship between hydrogen and current is 
\beq\label{eq:h-i}
h^{e}_{t} = k^e n^e I^e_{t}, 
\eeq
where we use a constant coefficient\footnote{Detailed expression can be found in the Appendix~\ref{sec:Hdetail}.} $k^e$, 
and $n^e$ is the number of electrolyzer cells connected in series.
% \tcr{which is linear because the amount of hydrogen produced is directly proportional to the amount of current flowing through the electrolyzer. For simplicity,
% The relationship can be further understood by Faraday's Law of Electrolysis {\cite[p.20]{bessarabov16pem}}
% Nonlinear V-I relation: $V^e(I^e)$, 
The input current for the PEM electrolyzer $I^e_t$ and the output power $p^e_t (kW)$ have a nonlinear relationship, given by
\beq\label{eq:p-i}
p^e_t =P_e(I^e_{t} n^e).
\eeq
Details about this nonlinear function $P_e(\cdot)$ are given in Appendix~\ref{sec:Hdetail}. Part of the nonlinearity is caused by the fact that energy is computed by the product of current and voltage. Another part of the nonlinearity comes from the fact that the PEM electrolyzer has nonlinear voltage-current characteristics, resulting from different voltage losses, e.g. activation losses, ohmic losses, and mass transport losses within the cell. The electrochemical kinetics within the cell are described by the Tafel Equation which has logarithmic terms making the polarization curves nonlinear{\cite[p.29]{bessarabov16pem}}. 

% \tcr{Cong: The writing above looks nice. For the following, what's the meaning of "one-and onto"? You can add a footnote explaining the definition of one-to-one function, or you can point to the appendix for readers who want to know the details of the models and parameters.}

Since $P_e(\cdot)$ is a bijective function, we get the following equation explaining the relationship between the input power $p^e_{t}$ and output hydrogen $h^e_{t}$ for the PEM electrolyzer by combining equation \eqref{eq:p-i} and \eqref{eq:h-i},
\beq\label{eq:he-pe}
h^e_t := f^e(p^e_t) =k^e[P_e]^{-1}(p^e_t).
\eeq
Some papers approximate this nonlinear relationship with $h^e_t \approx  (\eta^e  p^e_t)/{\cal H}$, where $\cal{H}$ $(kJ/kg)$ describes the energy content of hydrogen in terms of its higher heating value, and $\eta^e$ represents a constant efficiency parameter of a PEM electrolyzer, typically lying in the 70-90\% range{\cite[p.4]{bessarabov16pem}}. However, the performance comparison in our simulation demonstrates the inaccuracy of such a constant efficiency model. %To show the comparison with the constant efficiency parameter ($\eta^e$) is set to 80\%.

Due to high mass transport losses at high current densities, there is a limit on the maximum allowable current through the cell which thus imposes an upper bound on $p^e_{t}$ given by
\beq\label{eq:pe_lim}
0 \le p^e_t \le \bar{p}^e_{t}.
\eeq
 
\vspace{-0.5cm}
\subsection{Electrochemical model of  fuel cell}
% \tcr{Cong: 1) I marked red the sentence with grammar errors. 2) don't use abbreviations that you didn't explain the meanings of full names.}

The reverse reaction of using hydrogen to generate power is performed by a PEM fuel cell. The PEM fuel cell (PEMFC) technology is widespread commercially \cite{larminie03fuel}. In this study, we adopted the model from Corr\^ea et al. which was experimentally validated on a Ballard-Mark-V PEMFC\cite{correa2004electrochemical}. 

Similar to the electrolyzer, the linear relationship between the output current $I^f_{t} (A)$ and input hydrogen $h^f_{t} (kg/s)$\footnote{Superscript `$f$' represents fuel cell rather than power of a number.} is described by 
\beq\label{eq:hf-if}
I^{f}_{t} =k^f h^f_{t}/n^f,
\eeq
where $n^f$ represents the number of PEM fuel cells connected in series and $k^f$ is the constant coefficient that converts chemical input to electrical output.

% \tcr{Cong Mark 3: (7) is incorrect. double check derivation from (5) (6) to (7).}

The voltage-current relationships for a single PEMFC follow similar physics as an electrolyzer except that the reactions occuring at the anode and cathode are reversed\footnote{The detailed electrochemical model can be found in the Appendix~\ref{sec:Hdetail}.}. This results in a nonlinear relationship between the current generated by the fuel cell output power, $p^{f}_{t} (kW)$, given by the function
%\footnote{It is a bijective function when \eqref{eq:pf_lim} is satisfied.},
\beq\label{eq:pf-if}
p^{f}_{t} = P_f(I^f_{t}n^f).
\eeq
Details about this nonlinear function $P_f(\cdot)$ formula are explained in Appendix~\ref{sec:Hdetail}. 
Combining \eqref{eq:hf-if} and \eqref{eq:pf-if}  the relationship between the input hydrogen $h^f_{t}$ and output power $p^f_{t}$ for PEMFC is given by
\beq\label{eq:hf-pf}
p^f_t := f^f(h^f_t) = P_f(h^f_t k^f).
\eeq
% \tcr{$p^f_t := f^f(h^f_t) = P_f(h^f_t k^f).$}
Some papers approximate this nonlinear relationship \eqref{eq:hf-pf} with $p^f_t  \approx \eta^f  h^f_t {\cal H} $. The efficiency $\eta^f$ of a PEMFC typically lies in the 40-60\% \cite{eriksson2017optimization} range. %We set the constant efficiency parameter $\eta^f$ to 50\% in our simulation and verify the inaccuracy of such a constant efficiency model.

The upper bound on the power output of the fuel cell comes from the limit on the current through the cell, \ie %and is given by,
\beq\label{eq:pf_lim}
0 \le p^f_t \le \bar{p}^f_{t}.
\eeq

\vspace{-0.4cm}
\subsection{Intertemporal relationship of hydrogen storage}

The state of hydrogen (SoH) is the mass of hydrogen stored in the hydrogen tank, and it evolves with 
\begin{subequations}\label{eq:h_soc}
\begin{align}
&e_{t} + (h^e_t  - h^f_{t} ) \delta =e_{t+1},\label{eq:h_soc1}\\
&\underline{e} \le e_{t}\le \overline{e}, ~~e_{0} = 0.1\overline{e}.\label{eq:h_soc2}
\end{align}
\end{subequations}
where $e_{t} (kg)$ is the SoH  at time $t$ and $\delta$ is the time duration. Expressing SoH in $kg$ allows us to express the intertemporal relationship of the hydrogen storage as (\ref{eq:h_soc1}) without having to track the operating conditions (e.g. pressure and temperature) of electrolyzers, storage tank, and fuel cells separately. In this study, we assume the initial SoH to be 10\% in (\ref{eq:h_soc2}).

\vspace{-0.4cm}
\subsection{Piecewise linear approximation
}

% , \ie $h^e_t = \hat{f}^e(p^e_t)$ and $p^f_t =  \hat{f}^f(h^f_t)$,

We approximate the nonlinear electrochemical hydrogen storage model \eqref{eq:he-pe} and \eqref{eq:hf-pf} with piecewise linear functions,\ie  
%\begin{subequations}\label{eq:pe_linear-he_linear}
% \begin{align}
\beq\label{eq:pe_linear-he_linear}
\begin{array}{l}
h^e_{t} =\sum_{k=1}^K(a^e_k p^e_{t} + b^e_k)\mathbbm{1}\{p^e_{t} \in \Pc^e_k\},\\
% \beq\label{eq:hf_linear-pf_linear}
p^f_{t} =\sum_{k=1}^K(a^f_k h^f_{t} + b^f_k)\mathbbm{1}\{h^f_{t} \in \Hc^f_k\}.
% &\sum_{k=1}^K p^e_{t} = p^e_{t},~~\sum_{k=1}^K h^f_{t} = h^f_{t}.
\end{array}
\eeq
% \end{align}
% \end{subequations}
Here, $\Pc^e_k$ and $\Hc^f_k$ represent the $k$-th pieces for power and hydrogen flow, respectively. The indicator function $\mathbbm{1}$ means that $\mathbbm{1}(\Amsc)=1$ if $\Amsc$ is true and $\mathbbm{1}(\Amsc)=0$ otherwise. 
% Indicator functions  are used to denote membership of a point in a particular piece i.e. $\mathbbm{1}\{p^e_{t} \in \Pc^e_k\} = 1$ if $p^e_{t} \in \Pc^e_k$; and $\mathbbm{1}\{p^e_{t} \in \Pc^e_k\} = 0$ if $p^e_{t} \notin \Pc^e_k$. 
The coefficients of each linear piece $k$ are $a^e_k, b^e_k, a^f_k, b^f_k$, which is computed via minimizing the difference between \eqref{eq:he-pe},\eqref{eq:hf-pf}, and \eqref{eq:pe_linear-he_linear} for the piecewise linear hydrogen storage model.
%The detailed piecewise linear approximation model is in the appendix.
 % \tcr{make the plot below directly a H-P function.}

%  \begin{figure}[!htb]
%    \centering
%    \vspace{-0.1in}
%    \includegraphics[scale=0.37]{Fig/ez_ph.eps} \includegraphics[scale=0.37]{Fig/fc_ph.eps}
% \caption{\scriptsize  Piecewise linearized power-hydrogen relationship for a single cell. left: Electrolyzer; right: Fuel Cell. }
% \label{fig:PiecesApprox}
% \end{figure}

% \tcr{Fig.1 illustrates a piecewise linearization for the nonlinear hydrogen-power relationships from equations \eqref{eq:he-pe} and \eqref{eq:hf-pf}. Here, $K=2$ for the electrolyzer and $K=3$ for the fuel cell. From \eqref{eq:he-pe} and \eqref{eq:hf-pf}, it should be noted that these relationships were developed for a single PEM electrolyzer/fuel cell and their application to a stack comprising of cells connected in series can be extended by multiplying both $X$ and $Y$ axes by the number of cells $n^e$ or $n^f$. }

% \tcr{pay attention to the typical range of the x and y axis. Cite the reference to the parameters. explain the electrolyzer in cell and in series }

\vspace{-0.2cm}
\section{Energy management for microgrid resilience}\label{sec:AccessRight}
We first introduce the microgrid EMS with green hydrogen storage. Then we explain MPC together with the infeasibility correction method toward grid resilience in real-time.
\vspace{-0.4cm}
\subsection{ A microgrid model}

 \begin{figure}[!htb]
   \centering
   \vspace{-0.1in}
\includegraphics[scale=0.55]{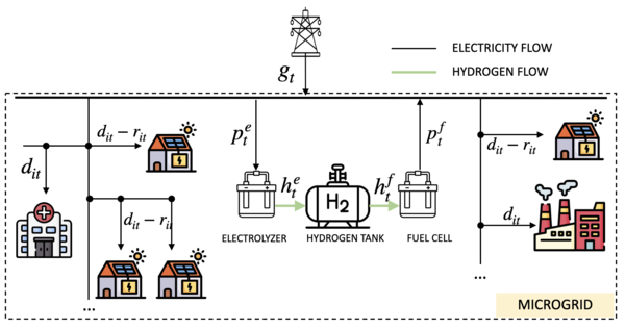}
\caption{\scriptsize  System schematics for the microgrid with hydrogen storage. }
\label{fig:Sys}
\end{figure}
\vspace{-0.2cm}

We adopt a simplified system schematic shown in Fig.~\ref{fig:Sys} for the microgrid energy management to enhance resilience with hydrogen storage. At time $t$, there are two electricity supply sources: the grid energy supply $g_t$ and the roof-top solar $r_{it}$ from household $i$. Some of this electricity is directly consumed by the household, denoted by $d_{it}$, while another portion is used to produce hydrogen through electrolyzers, which is subsequently stored in a hydrogen tank. During contingencies, the hydrogen in the tank can be utilized to generate supplementary electricity using fuel cells.    

\vspace{-0.3cm}
\subsection{Resilence-enhancing microgrid EMS}
We considered a microgrid containing customers with different values of the lost load  $(v_1,...,v_N)$.  $N$ is the number of households. Critical loads like hospitals have higher values of the lost load.  The loss-of-load for customer $i$ at time $t$ is denoted by $l_{it}$. The electricity price from the grid supply is $c_t \in \mathbb{R}_+$. The objective of EMS is to minimize the penalty induced by the loss-of-load and the cost of purchasing energy from the grid. We consider three types of loss-of-load penalties: 
\begin{itemize}
    \item Type 1 (eg. \cite{yao2022quantitative}) adopts $l_{1}$-norm $\Phi_i(\lbf_i) = v_i||\lbf_i||_1 $ to penalize  the total loss-of-load.
    \item Type 2 (eg. \cite{hussain2017resilience}) adopts $l_{2}$-norm $\Phi_i(\lbf_i) = v_i||\lbf_i||_2 $ to give more penalty for the large value of loss-of-load. 
    \item  Type 3 (we propose) adopts a mixed norm of $l_{1}$ and $l_{\infty}$ norms  $\Phi_i(\lbf_i) = v_i||\lbf_i||_1 + v_i||\lbf_i||_\infty \cdot T$ to penalize both the total loss-of-load and maximum load shed.%\footnote{Note that if we only adopt the $l_{\infty}$-norm for the penalty function $\Phi_i(\lbf_i) = v_i||\lbf_i||_\infty \cdot T$, the microgrid tends to maintain a constant load shedding level all the time even after the grid supply recovered.} %The intuition of this penalty function is to use proper weights to construct the weighted sum of two optimization objectives.
% \tcr{\item Type 4 adopts proportion fairness and replace $\sum_{i=1}^N\Phi_i(\lbf_i)$ in the objective with penalty function $\Phi(\lbf_1,...,\lbf_N) = \sum_{i=1}^N log(v_i ||\lbf_i||_1)$. Details about proportion fairness are explained in Sec.~\ref{sec:Fairness} of the appendix.
% }

% \tcr{Cong: if you have time to try this, the fairness for designing the penalty function would be a nice new contribution. Previously Prof.Tong recommended the fairness consideration to us, and I find this simple model to implement. 1) It's a slight modification  of the Type 1 penalty, and you need to pay attention that this function is not well-defined when $||\lbf_i||_1=0$. If you meet this case, try to make it piecewise as what you're implemented now for the nonlinear model which is also not well-defined with $log()$ function. 2) You probably need to make the hospital have a very large penalty $v_i$ compared to the household $v_i$, you can try penalty 10, 1, 0.1 for those three groups you considered.}
\end{itemize}

The optimization for microgrid energy management is adopted in the receding horizon for MPC. In each receding horizon from the initial time  $t'$ to the end time $T$, the length of the time horizon is ${\cal T}:= T-t'+1$. The roof-top solar generation prediction $(r_{it})$ and the initial SoH $e_{t'}$ are given. Define  $\Omega := \{\pbf^f, \pbf^e, \hbf^f, \hbf^e, \ebf, \gbf \in \mathbb{R}_+^{\cal T} , \Dbf, \Lbf \in \mathbb{R}_+^{N\times {\cal T}} \}$ for the domain of  decision variables, $\Dbf:=(d_{it})$, and $\Lbf:=(l_{it})$.  The resilience-enhancing microgrid EMS is given by
\vspace{-0.2cm}
\begin{subequations} 
    \begin{align}
    &  \underset{\Omega}{\rm minimize} &&  \sum_{i=1}^N\Phi_i(\lbf_i)+\sum_{t=t'}^{T} c_t g_t, \label{eq:obj}
    \\
    & \text{subject to} 
    && \forall t\in \{t',..., T\}, \forall i\in \{1,..., N\} \nn 
    \\
    % &&& H_{t} + (f^e(p^e_t)  - f^f(p^f_t) ) \delta =H_{t+1},
    % &&& \sum_{i=1}^N (e_{it} - l_{it})- p^f_t - g_t + p_t^e = 0,
    % \\
    &&& d_{it} - r_{it}
 - l_{it} =q_{it},\label{eq:q1}\\
    &&& p^e_t - p^f_t - g_t = q_{0t},\label{eq:q2}\\ 
    &&& \ul{\v{b}} \leq \v{A} \qbf_t \leq \ol{\v{b}},\label{eq:distflow}\\
    &&& \l_{it} \le d_{it},~~ g_t \le \bar{g}_t, \label{eq:g_const} \\&&&\underline{d}_{it} \le d_{it}\le \bar{d}_{it} \label{eq:dd_const},\\
    &&&\eqref{eq:pe_lim},\eqref{eq:pf_lim},\eqref{eq:h_soc},\eqref{eq:pe_linear-he_linear}.\nn
    \end{align}
    \label{eq:EM_multi}
\end{subequations}
  For simplicity, we assume each bus only has one customer. In \eqref{eq:q1}\eqref{eq:q2}, $q_{it}$ is the net withdraw power from but $i$ at time $t$. We have $\qbf_t \in \Rset^{M}$ for a $M$-bus  network. We assume the hydrogen storage is connected to the slack bus, the grid connection point in the network. In \eqref{eq:distflow}, we use a linearized DistFlow (LinDistFlow) model for the microgrid network\cite{ChenBoseTong22Access}. $\Abf \in \Rset^{N \times M}$ is the parameter for the LinDistFlow constructed by the network impedance, resistance, and topology information. $\ol{\v{b}}, \ul{\v{b}}$ are composed of network voltage limits and thermal limits. Details about parameters $\Abf, \ol{\v{b}}, \ul{\v{b}}$ are in the appendix of \cite{ChenBoseTong22Access}. \eqref{eq:g_const} shows that the customer loss-of-load  $l_{it}$ is constrained by the household demand, and the grid supply $g_{t}$ is constrained by $\bar{g}_{t}$, the maximum available grid supply. In \eqref{eq:dd_const}, $\underline{d}_{it}$ and $\bar{d}_{it}$ are boundary limits on customer  demand. 

  The piecewise linear approximation \eqref{eq:pe_linear-he_linear}  of hydrogen storage is adopted in this optimization to substitute nonlinear equality constraints \eqref{eq:he-pe} and \eqref{eq:hf-pf}, which will result in a nonconvex optimization. Thus, we reduce the computation burden while maintaining an accurate solution. We further formulate (\ref{eq:EM_multi})  into a mixed integer program (MIP) with linear constraints and convex objective (details in Appendix~\ref{sec:MILP}).

\vspace{-0.3cm}
\subsection{Model predictive control with feasibility projection}

The real-time microgrid operation is conducted in a receding window MPC.  In both the Type I linear model and the Type II piecewise linear model for storage, the optimal solution might be infeasible under the nonlinear hydrogen storage model. We propose a feasibility projection after finishing the optimization in each receding window. Such a feasibility projection can fix the approximation error in real-time. Here is a high-level summary for the feasibility projection of hydrogen storage operation. If infeasible hydrogen dispatch exceeds the maximum bounds $\overline{e}$, we decrease the electrolyzer output to make SoH bounded by $\overline{e}$; if infeasible SoH exceeds the minimum bounds $\underline{e}$, we decrease the fuel cell output correspondingly. After achieving a feasible operation for hydrogen storage, we rerun the optimization in that receding window by fixing the hydrogen operation (See detail algorithm in Appendix\ref{sec:FP}).

\vspace{-0.1cm}
\section{Numerical Results}\label{sec:SOAccessRight}
\vspace{-0.0015cm}
We conducted simulations comparing three hydrogen storage models with different loss-of-load penalty functions. (i) The linear model adopted $h^e_t = (\eta^e p^e_t)/{\cal H}, p^f_t = \eta^f h^f_t {\cal H} $ with $\eta^e = 80\%$ and $\eta^f = 50\% $ for hydrogen storage via receding window MPC. (ii) The proposed piecewise linear storage model was implemented as \eqref{eq:EM_multi} via receding window MPC. (iii) The nonlinear model \eqref{eq:he-pe} and \eqref{eq:hf-pf} was simulated as an ideal benchmark, which adopted the perfect forecast and solved the dispatch in one shot. The Appendix~\ref{sec:sim} contains parameters of detailed nonlinear models and additional results. 
   \vspace{-0.2in}
 \subsection{Parameter settings}
We simulated a 60-min time horizon with a 40-min island mode contingency for a hypothetical microgrid\footnote{The network constraints were set to be unbinding for the microgrid, equivalent to a single bus system.} with 20 households. We considered three categories of customers with different values of lost load\footnote{Values of lost load and pricing parameters were set based on   \cite{ERCOT23ScarcityPrice}.}, \ie type-1  $\vbf_1 = \$5$/kWh, type-2 $\vbf_2 = \$1$/kWh, and  type-3 $\vbf_3 = \$0.5$/kWh. There were 5,7, and 8 customers in each of these categories. Demand lower bounds ${\underline{\dbf}}$ were set to be 13, 10, and 8 kW respectively for the three categories. We set $c_{t} = \$ 0.1$/kW, $\forall t$  for the outer grid electricity price, the maximum hydrogen storage charge and discharge rates 150 kW ($\bar{p}^e$) and 70 kW ($\bar{p}^f$), the storage tank capacity ranged from 0 to 3 kg ($\underline{e}$ and $\overline{e}$), the higher heating value of hydrogen 142$MJ/kg$ ($\cal{H}$).
%$k^e=$ 0.0027 $kg\cdot cm^2/C$, and $ k^f =$ 1316.5 $C/cm^2\cdot kg$. 

% \subsection{Stochastic scenarios}

% \tcr{Cong: 1) what's the unit for 1200, 100, 1...? 2) finish the the whole sentence revision based on the beginning I provided.Shreshtha:Updated}

To simulate the 40-min island mode contingency, the maximum grid supply capacity $\bar{g}_t$ was 0 kW (\ie complete outage) for $t \in \{11,..,40\}$, and 1000 kW for other intervals. We used Pecan Street data\footnote{The data is accessible at \href{https://www.pecanstreet.org/dataport/}{Pecan St. Project}.} for household consumption and rooftop solar generation. Solar forecasts $r_{it}$, took the mean data from 7 PM - 8 PM with Gaussian noise, $\mathcal{N}(0,1)$ kW,  added. 
% was sampled and was considered as a deterministic case (assuming the renewable completely known) to solve the non-linear model optimization which was solved in one-shot. For the MPC forecast module, we generate a new realization of $\hat{r}_{t:T}$ for each receding horizon at time $t$ from $t+1$ to $T$ and use the deterministic value at time $t$ from ${r}^{'}_{t}$. 

\vspace{-0.1in}
\subsection{Performance evaluation}

A resilient microgrid anticipates small total loss-of-load and large minimum system load\footnote{This refers to the minimum of the resilience trapezoid in Fig.~\ref{fig:trapezoid}, If the maximum loss-of-load increases, then the minimum system load \% decreases.} the microgrid can sustain during the contingency. We define the dominated area for one marker as its top-left area in Fig.~\ref{fig:pareto} because points in the dominated area perform worse than the markers for both resilience metrics. 
 \begin{figure}[!htbp]
   \centering
   \vspace{-0.1in}
 \includegraphics[width=0.28\textwidth]{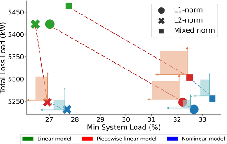}   %\includegraphics[scale=0.15]{Fig/ParetoC1.png}    %\includegraphics[scale=0.15]{Fig/ParetoC2.png}
\caption{\scriptsize  Total loss-of-load of the microgrid and the minimum system load sustained. (Top left of each marker is the dominated area.)}
\label{fig:pareto}
\end{figure}

Our observations for these two resilience metrics in Fig.~\ref{fig:pareto} are three-fold. First, the linear model is always dominated by the piecewise linear model, because the green markers always fall into the dominated area of the red markers under different penalty functions. Second, the piecewise linear model approaches the nonlinear model, the ideal benchmark, from which the linear model is far away. This is represented by the length of dash curves in Fig.~\ref{fig:pareto}, and we observe that the red curves are longer than the blue. Third, comparing different penalty functions, the $l_{1}$-norm penalty (circular markers) consistently results in the lowest total loss-of-load, attributed to the direct penalization for this metric; the mixed-norm penalty (square markers) outperforms in the minimum system load due to its $l_{\infty}$-norm penalizing extreme load-shedding.
 \vspace{-0.15in}
\begin{table}[!htbp]
\caption{System cost}
\label{tab:SC}
\centering
\begin{tabular}{@{}ccccccc@{}}
\toprule
\multirow{2}{*}{Penalty} & \multicolumn{2}{c}{Linear} & \multicolumn{2}{c}{Piecewise Linear}  & \multicolumn{2}{c}{Nonlinear}                \\ \cmidrule(l){2-7} 
                      & Cost       & \% +/-   & Cost   & \% +/- & Cost & \% +/-  \\ \midrule
$l_{1}$-norm & \$5127 & +17\% & \$4415 & +0.8\%& \$4377 & -\\ 
$l_{2}$-norm & \$5956 & +6.4\% & \$5614 & +0.3\% & \$5596 & -\\
Mixed norm & \$5361 & +21.7\% & \$4448 & +0.9\% & \$4404 & -\\ \bottomrule
\end{tabular}
% \vspace{-0.3in}
\end{table}
System costs, which include the value of lost load and the cost of grid energy supply (computed by $\sum_{t=0}^{T-1} (v_i l_{it} + c_t g_t)$) are shown in Table~\ref{tab:SC}. The costs of linear and piecewise linear storage models are expressed as a percentage increase/decrease relative to the cost of the ideal nonlinear benchmark in the column ``\% +/-".  Two key observations emerge. First, there exists a less than 1\% disparity in cost between the piecewise linear and nonlinear models. In contrast, the linear model results in a 6.4\%-21.7\% increase in cost compared to the benchmark. Second, the lowest cost is attained with the $l_{1}$-norm penalty, given its direct minimization within the objective function.
% \vspace{-0.1in}
% \vspace{-0.1in}         
\begin{table}[!htbp]
\caption{Time metrics for type-2 customers}
\label{table:time_metrics}
\centering
\begin{tabular}{@{}ccccccc@{}}
\toprule
\multirow{2}{*}{Penalty} & \multicolumn{2}{c}{Linear} & \multicolumn{2}{c}{Piecewise Linear}  & \multicolumn{2}{c}{Nonlinear}                \\ \cmidrule(l){2-7} 
                      & DO       & BTO  & DO   & BTO & DO & BTO   \\ \midrule
$l_{1}$-norm                     & 51.67\%  & 11 & 38.34\% & 22 & 38.34\% & 22\\
$l_{2}$-norm                     & 23.34\%  & 21 & 20\% & 31 & 5\% & 41\\
Mixed norm                     & 50\% & 11 & 26.67\% & 21 & - & -\\ \bottomrule
\end{tabular}
% \vspace{-0.1in}
\end{table}
The metrics pertaining to time, i.e. the duration-of-outage (DO) as a percentage of the 60-min time horizon and the beginning time of outage (BTO), which corresponds to the time (in min) at which the system load percentage first goes below 1\% in Fig.~\ref{fig:trapezoid} (second row) are tabulated in Table~\ref{table:time_metrics}\footnote{`-' implies there is no outage in Table~\ref{table:time_metrics}.}. Note that for the DO we consider the total time for which system load percentage is below 1\%. From the data presented in Table~\ref{table:time_metrics}, two key observations emerge. First, as seen for all the other metrics, the linear model deviates significantly in performance from the ideal benchmark. On the other hand, the piecewise linear model closely approximates the ideal benchmark. 
Specifically, in comparison to the linear model, the piecewise linear model reduces the DO by 13\%-48\%. 
Second, when subject to the $l_{2}$-norm penalty, DO is reduced, and BTO is delayed, signifying that the outage takes place later in the 60-min time horizon. This pattern is consistent across all three storage models. We observe a 54.7\% reduction in DO for the linear model, while a rather significant 87\% %\footnote{This range was calculated based on the least to highest improvement amongst the different storage models.\label{footnote_6}} 
reduction for the nonlinear model. The BTO is delayed by 41\%-91\% under the $l_{2}$-norm penalty.\footnote{See Appendix~\ref{sec:sim} for more simulations exploring this phenomenon.}
%\footref{footnote_6}. 
% Additional simulation results to explore this phenomenon are included in the appendix. %While $l_{2}$-norm performs close to $l_{1}$-norm in total loss-of-load, as shown by the cross and circular markers in Fig.~\ref{fig:pareto}, it's essential to highlight the distinct manner in which loss-of-load is allocated to different customer categories under $l_{2}$-norm compared to $l_{1}$-norm and Mixed norm. In Fig.~\ref{fig:trapezoid} (first row, second column) it's evident that loss-of-load for type-1 customers under the $l_{2}$-norm penalty significantly exceeds that of $l_{1}$-norm and Mixed norm. Conversely, as shown in Fig.~\ref{fig:trapezoid} (second row, second column) the loss-of-load for type-2 customers under the $l_{2}$-norm penalty is reduced in comparison to $l_{1}$-norm and Mixed norm, resulting in a shorter duration-of-outage. This highlights $l_{2}$-norm's role in promoting a more equitable distribution of loss among different customer categories. 

\begin{figure}[!htbp]
   \centering
   \vspace{-0.1in}
  \includegraphics[width=0.48\textwidth]{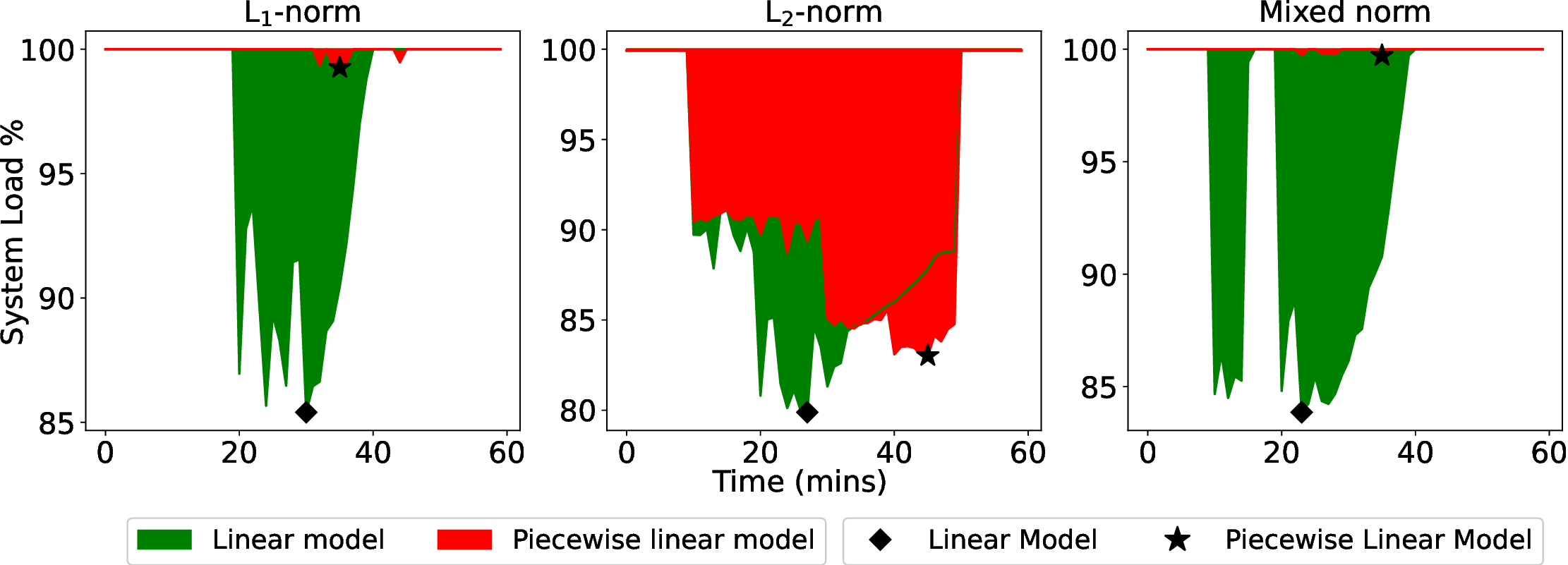}    \includegraphics[width=0.48\textwidth]{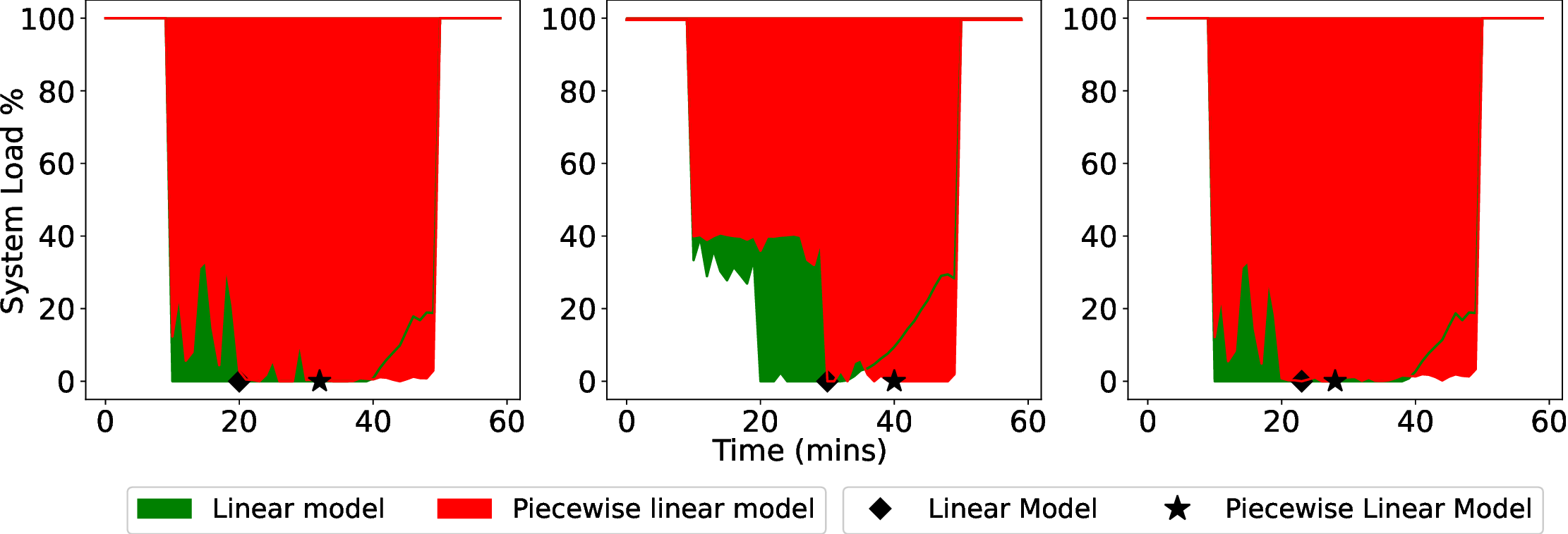}
\caption{\scriptsize  Top to Bottom:  Resilience trapezoids for type-1 customers and type-2 customers. Left to Right: Resilience trapezoids under different penalty schemes (The diamond and star markers represent the minimum points).}
\label{fig:trapezoid}
\end{figure}
In Fig.~\ref{fig:trapezoid}, we depict resilience trapezoids with total system load percentage, which is the load demand excluding loss-of-load as a percent of the total demand on the $Y$-axis. The ``resilience trapezoid"\cite{panteli2017metrics} provides a visualization of the system states. The shaded area represents the total loss-of-load percentage and the markers represent the minimum system load percentage. Comparing the linear and piecewise linear models we observe: (i)
For type-1 customers (Fig.~\ref{fig:trapezoid} first row), the piecewise linear model reduces total loss-of-load by 95\% under the $l_{1}$ and mixed-norm penalties and 12\% under the $l_{2}$-norm compared to the linear storage model; (ii) For type-2 customers (Fig.~\ref{fig:trapezoid} second row), reductions of 6\% under the $l_{2}$-norm and 2\% under the $l_{1}$-norm are observed.

% the following:
%The area enclosed by the $Y = 100$ line and the trapezoid is proportional to the {\em total loss-of-load}. The piecewise linear storage model (red) closely approaches the ideal benchmark (blue), deviating by 0.8\% (Fig.~\ref{fig:trapezoid} first row, third column) to 1.7\% (Fig.~\ref{fig:trapezoid} second row, third column) in terms of the total loss-of-load. 
% \footnote{This corresponds to the time (min) at which the system load \% first goes 0 in Fig.~\ref{fig:trapezoid} (second row).}
\vspace{-0.2cm}
\section{Conclusions}\label{sec:Conclusion}

% \tcr{Cong: You can try to write the conclusion with several sentences. And then connect it with the paragraph I wrote below about the limitations of this paper.}
To improve electricity supply resilience, we co-optimize microgrid and hydrogen storage operation with the proposed piecewise linearization model, capturing the nonlinear electrochemical model of electrolyzers and fuel cells. This approach  implemented with a feasibility projection method in model predictive control accurately approximates the performance of the ideal nonlinear benchmark model for hydrogen storage.  Besides, this paper has limitations in considering only operations during the contingency.  In our future work, we will explore joint optimization for both pre-contingency and during-contingency operations.%, and consider various events like the outage caused by the line tripping \cite{Guo21outage_I}.  

%The distribution of islanding mode frequency for the microgrid is not explored, and the detailed outer grid blackout scenarios are not considered.

\vspace{-0.2cm}
{
\bibliographystyle{IEEEtran}
\bibliography{BIB}

\begin{thebibliography}{10}
\providecommand{\url}[1]{#1}
\csname url@samestyle\endcsname
\providecommand{\newblock}{\relax}
\providecommand{\bibinfo}[2]{#2}
\providecommand{\BIBentrySTDinterwordspacing}{\spaceskip=0pt\relax}
\providecommand{\BIBentryALTinterwordstretchfactor}{4}
\providecommand{\BIBentryALTinterwordspacing}{\spaceskip=\fontdimen2\font plus
\BIBentryALTinterwordstretchfactor\fontdimen3\font minus
  \fontdimen4\font\relax}
\providecommand{\BIBforeignlanguage}[2]{{%
\expandafter\ifx\csname l@#1\endcsname\relax
\typeout{** WARNING: IEEEtran.bst: No hyphenation pattern has been}%
\typeout{** loaded for the language `#1'. Using the pattern for}%
\typeout{** the default language instead.}%
\else
\language=\csname l@#1\endcsname
\fi
#2}}
\providecommand{\BIBdecl}{\relax}
\BIBdecl

\bibitem{zhang2017comparative}
Y.~Zhang, P.~E. Campana, A.~Lundblad, and J.~Yan, ``Comparative study of
  hydrogen storage and battery storage in grid connected photovoltaic system:
  Storage sizing and rule-based operation,'' \emph{Applied energy}, vol. 201,
  pp. 397--411, 2017.

\bibitem{tobajas22AE}
J.~Tobajas, F.~Garcia-Torres, P.~Roncero-S{\'a}nchez, J.~V{\'a}zquez,
  L.~Bellatreche, and E.~Nieto, ``Resilience-oriented schedule of microgrids
  with hybrid energy storage system using model predictive control,''
  \emph{Applied Energy}, vol. 306, p. 118092, 2022.

\bibitem{haggi2022proactive}
H.~Haggi, W.~Sun, J.~M. Fenton, and P.~Brooker, ``Proactive
  rolling-horizon-based scheduling of hydrogen systems for resilient power
  grids,'' \emph{IEEE Transactions on Industry Applications}, vol.~58, no.~2,
  pp. 1737--1746, 2022.

\bibitem{Gabrielli16EEEIC}
P.~Gabrielli, B.~Flamm, A.~Eichler, M.~Gazzani, J.~Lygeros, and M.~Mazzotti,
  ``Modeling for optimal operation of pem fuel cells and electrolyzers,'' in
  \emph{2016 IEEE 16th International Conference on Environment and Electrical
  Engineering (EEEIC)}, 2016, pp. 1--7.

\bibitem{FLAMM21AE}
\BIBentryALTinterwordspacing
B.~Flamm, C.~Peter, F.~N. Büchi, and J.~Lygeros, ``Electrolyzer modeling and
  real-time control for optimized production of hydrogen gas,'' \emph{Applied
  Energy}, vol. 281, p. 116031, 2021. [Online]. Available:
  \url{https://www.sciencedirect.com/science/article/pii/S0306261920314690}
\BIBentrySTDinterwordspacing

\bibitem{bessarabov16pem}
D.~Bessarabov, H.~Wang, H.~Li, and N.~Zhao, \emph{PEM electrolysis for hydrogen
  production: principles and applications}.\hskip 1em plus 0.5em minus
  0.4em\relax CRC press, 2016.

\bibitem{larminie03fuel}
J.~Larminie, A.~Dicks, and M.~S. McDonald, \emph{Fuel cell systems
  explained}.\hskip 1em plus 0.5em minus 0.4em\relax J. Wiley Chichester, UK,
  2003, vol.~2.

\bibitem{abomazid21TII}
A.~M. Abomazid, N.~A. El-Taweel, and H.~E. Farag, ``Novel analytical approach
  for parameters identification of \uppercase{PEM} electrolyzer,'' \emph{IEEE
  Transactions on Industrial Informatics}, vol.~18, no.~9, pp. 5870--5881,
  2021.

\bibitem{correa2004electrochemical}
J.~M. Corr{\^e}a, F.~A. Farret, L.~N. Canha, and M.~G. Simoes, ``An
  electrochemical-based fuel-cell model suitable for electrical engineering
  automation approach,'' \emph{IEEE Transactions on industrial electronics},
  vol.~51, no.~5, pp. 1103--1112, 2004.

\bibitem{yao2022quantitative}
Y.~Yao, W.~Liu, R.~Jain, B.~Chowdhury, J.~Wang, and R.~Cox, ``Quantitative
  metrics for grid resilience evaluation and optimization,'' \emph{IEEE
  Transactions on Sustainable Energy}, vol.~14, no.~2, pp. 1244--1258, 2022.

\bibitem{NAN201735}
\BIBentryALTinterwordspacing
C.~Nan and G.~Sansavini, ``A quantitative method for assessing resilience of
  interdependent infrastructures,'' \emph{Reliability Engineering \& System
  Safety}, vol. 157, pp. 35--53, 2017. [Online]. Available:
  \url{https://www.sciencedirect.com/science/article/pii/S095183201630374X}
\BIBentrySTDinterwordspacing

\bibitem{panteli2017metrics}
M.~Panteli, P.~Mancarella, D.~N. Trakas, E.~Kyriakides, and N.~D.
  Hatziargyriou, ``Metrics and quantification of operational and infrastructure
  resilience in power systems,'' \emph{IEEE Transactions on Power Systems},
  vol.~32, no.~6, pp. 4732--4742, 2017.

\bibitem{hussain2017resilience}
A.~Hussain, V.-H. Bui, and H.-M. Kim, ``Resilience-oriented optimal operation
  of networked hybrid microgrids,'' \emph{IEEE Transactions on Smart Grid},
  vol.~10, no.~1, pp. 204--215, 2017.

\bibitem{eriksson2017optimization}
E.~Eriksson and E.~M. Gray, ``Optimization and integration of hybrid renewable
  energy hydrogen fuel cell energy systems--a critical review,'' \emph{Applied
  energy}, vol. 202, pp. 348--364, 2017.

\bibitem{ChenBoseTong22Access}
C.~Chen, S.~Bose, and L.~Tong, ``{DSO}-{DERA} coordination for the wholesale
  market participation of distributed energy resources,'' \emph{arXiv
  preprint}, 2022.

\bibitem{ERCOT23ScarcityPrice}
``{ERCOT} staff recommendation to committee,'' [ONLINE], available at
  \url{https://www.ercot.com/files/docs/2023/04/11/10.1%20Phase%202%20Market%20Redesign%20-%20Bridging%20Solutions.pdf},
  April 2023.

\end{thebibliography}
}

% \newpage
\section{Appendix}
\label{sec:Appendix}
% \subsection{Format to use}

% % \tcr{other table format}
% % {\small 
% % \begin{table}[]
% % \caption{Surplus distribution for the 4-bus example}
% % \label{table:surplus.4-bus}
% % \centering
% % \begin{tabular}{llllll}\toprule
% % \centering
% % Allocation &  DERA$_1$ & DSO & DERA$_2$ & Social Surplus \\
% % \midrule
% % Robust  & 198.25 & 282.6  & 748.25 & 1229.1 \\ 
% % Stoch ($\delta = 0.90)$  & 209.51 & 404.36   & 759.61 & 1373.48 \\ \bottomrule
% % \end{tabular}
% % \end{table}
% % }

% \tcr{other table format}
%      \begin{center}
%         \begin{tabular}{ c|c|c|c } 
%          \hline
%           & $h_1$ correct & $h_1$ incorrect & Row total\\ 
%          \hline
%          $h_2$ correct & 240 & 9 & 249\\ 
%          $h_2$ incorrect & 15 & 21 & 36\\ 
%          Column total & 255 & 30 & 285\\
%          \hline
%         \end{tabular}
%         \end{center}

% \tcr{here is a table format}
% \begin{table}[]
% \caption{Variation of DERA surplus with $\sigma$}
% \label{table:DERASS}
% \centering
% \begin{tabular}{@{}ccccc@{}}
% \toprule
% \multirow{2}{*}{DERA} & \multicolumn{4}{c}{$\sigma$ (MW)}       \\ \cmidrule(l){2-5} 
%                       & 0       & 0.004   & 0.006   & 0.008   \\ \midrule
% 1                     & 599.54  & 488.00  & 431.20  & 369.41  \\
% 2                     & 324.07  & 291.43  & 277.58  & 265.01  \\
% 3                     & 1043.85 & 1042.54 & 1042.41 & 1042.41 \\
% 4                     & 80.18   & 76.74   & 75.09   & 73.49   \\ \bottomrule
% \end{tabular}
% \vspace{-0.1in}
% \end{table}

\subsection{Detailed electrochemical model of hydrogen storage}\label{sec:Hdetail}

% \tcr{Cong: could you write in this way $T= 353$ K rather than $T(K)= 353$ when introduing values for the parameters?}

We adopted the model for PEM electrolyzer from \cite{abomazid21TII} and the Ballard-Mark-V PEMFC model from \cite{correa2004electrochemical}. In this section, we drop the time index subscript `$t$' for simplicity.
% \tcr{Cong: 1) could you cite the paper you refer to for the model below, before you introduce details? You can say "We refer to [X] for the nonlinear hydrogen storage model ..." 2) I mark red some typos or grammar errors. 3) try $ln(p_{H_{2}})$ rather than $lnp_{H_{2}}$. 4) could you explain the value you take for the parameters? rather than let readers read another paper to understand the parameter you choose. 5) In equation (17), the unit of $r$ is strange. Also you didn't explain $J^e$ in that equation. Move that under equation (18) to equation (17). 6) you didn't explain $J^e_L$ in (21), and also the value you use in the simulation. 7) use $\bar{I}^f$ rather than $\tcr{\bar{I^f}}$. 8) I directly modify equation Sec.VI-B, if you don't agree with anything, feel free to do revisions.}

% \subsubsection{Electrochemical model of electorlyzer $h^e_t = f^e(p^e_t)$}

% \subsubsubssection
\subsubsection{Electrochemical model for electrolyzer $h^e := f^e(p^e)$}
\begin{itemize}
\item{Hydrogen-Current Relationship}

\beq\label{eq:h-I_relation}
h^e = n^e \frac{  I^e  T  R d_{H_{2}}}{2F  p_{H_{2}}}
\eeq

Here, $h^e(kg/s)$ is the amount of hydrogen produced, $I^e(A)$ is the input current to the electrolyzer, $n^e$ is the number of electrolyzer cells in series, $T (K)$ is the operating temperature of the electrolyzer cells, $p_{H_{2}} (bar)$ is the hydrogen partial pressure, $d_{H_{2}} (kg/m^3)$ is the density of hydrogen at the operating conditions, $F (C/mol)$ is Faraday's constant, and $R (m^3 \cdot bar \cdot K^{-1} \cdot mol^{-1})$ is the ideal gas constant. In the modeling section, we express \eqref{eq:h-I_relation} as $h^e = k^en^eI^e$ in \eqref{eq:h-i}. We therefore have,
\beq
k^e = \frac{ T  R d_{H_{2}}}{2F  p_{H_{2}}}.
\eeq
In the simulation section, we set $n^e=150$, $T= 353$ $K$, $p_{H_{2}}= 1$ $bar$, $F = 96485$ $C/mol$, $d_{H_{2}}= 0.07$ $kg/m^3$, $R = 8.314 \cross 10^{-5}$ $m^3 \cdot bar \cdot K^{-1} \cdot mol^{-1}$ and $k^e= 0.0027$ $kg\cdot cm^2/C$.
 
\item{Voltage-Current Relationship}
\newline
The voltage-current relationship for a single PEM electrolyzer cell can be expressed as
\beq\label{eq:vEZ_1}
V_{e}(I^e) = V^e_{Nernst} + V^e_{act} + V^e_{\Omega} + V^e_{con},
\eeq

where $V^e_{Nernst}$ denotes the open circuit voltage, $V^e_{\Omega}$ the ohmic overvoltage, $V^e_{act}$ the activation overvoltage, and $V^e_{con}$ the concentration overvoltage. It should be noted that when $I^e = 0$ i.e. there is no current through the cell, $V_{e}(I^e)$ is simply equal to the open circuit voltage, $V^e_{Nernst}$ which is given by

\beq\label{h-I relation}
V^e_{Nernst} = \frac{\Delta G}{2F} + \frac{RT}{2F}ln\bigg(\frac{p_{H_{2}}\sqrt p_{O_{2}}}{p_{H_{2}O}}\bigg),
\eeq

where $\Delta G (kJ/mol)$ is the change in Gibbs free energy, $p_{O_{2}}(bar)$, $p_{H_{2}O}(bar)$ are the partial pressures of oxygen and water. The change in Gibbs free energy denotes the minimal amount of free energy required to form one mole of a substance from its inert component. The computation of the Gibbs free energy takes into account the effect of various temperatures and pressures on the electrolysis process. In the simulation section, we set $\Delta G =233.102$ $kJ/mol$, $p_{O_{2}}= 0.5$ $bar$, and $p_{H_{2}O}=0.5$ $bar$.
\newline
The activation overvoltage $V^e_{act}$ is created as a result of the movement of protons and electrons between the cathode and anode in the electrochemical reaction and is given by
\beq\label{h-I relation}
V^e_{act} = \frac{RT}{2F\alpha} arcsinh\bigg(\frac{J^e}{2J_{0}}\bigg),
\eeq
\beq
J^e = \frac{I^e}{A^e},
\eeq
where $\alpha$ is the charge transfer rate, $J^{e} (A/cm^2)$ is the current density through the cell, $J_0 (A/cm^2)$ is the exchange current density and $A^e (cm^2)$ is the active surface area of the cell.  In the simulation section, we set $\alpha =0.2993$ and $J_0  = 13.4776 \cross 10^{-6}$ $A/cm^2$, and $A^e = 160$ $cm^2$.
\newline 
The ohmic losses are described by
\beq\label{h-I relation}
V^e_{\Omega} = J^er,
\eeq
where $r (\Omega \cdot cm^2)$ is the total internal resistance of the PEM electrolyzer cell. In the simulation section, we set $r =0.2614$ $\Omega\cdot cm^2$.
\newline
Finally, $V^e_{con}$ describes the overvoltage caused at high current densities\footnote{Due to mass transport resistance.} and can be expressed as
\beq\label{eq:vEZ_2}
V^e_{con} = \frac{RT}{2F}ln\bigg(\frac{J^e_{max}}{ J^e_{max} -J^e}\bigg),
\eeq
where $J^e_{max} (A/cm^2)$ is the limiting current density, which is the maximum allowable current density for electrolysis. In the simulation section, we set $J^e_{max}  =2.146$ $A/cm^2$.
% $\color{blue} h^{e}(I^e) \color {black} = N^e \frac{\color{blue} I^e \color{black}\cdot T \cdot R}{2F \cdot p_{H_{2}}}$ 
% \newline
% this can be simplified to:
% \newline
% $ h^{e}(I^e)  =  K^e \cdot I^e $ 

% V-I relation:
% $V^e = f_{vi}(I^e)$, 

\item{Power-Current Relationship}
\newline 
For $n^e$ cells in series, the power-current relationship is described by
\beq\label{eq:p-i_ez}
P_{e}(I^e) = n^e I^e V_{e}(I^e).
\eeq
where, $V_{e}(I^e)$ is described by equations \eqref{eq:vEZ_1}-\eqref{eq:vEZ_2}. When $I^e = 0$, the cell is in an open circuit condition, implying $P_{e}(0) = 0$.
% \tcr{Cong: You need to connect $V^e(I^e)$ with equations (15)-(19), otherwise people don't know how those equations are related to each other. I guess you need to write $J^e$ the current density equals $I^e$ divided by the area. Also, you need to explain the parameter of the area you choose in your simulation. I guess you need to use $I^e=A^eJ^e$.}

\item{Operation Limits}

The operation limit in a single PEM electrolyzer cell is given by
\newline
\beq\label{eq:i_ez_lim}
0 \le I^e \le \bar{I^e}:=A^e J^e_{max}.
\eeq
% \end
We can use equations \eqref{eq:p-i_ez} and \eqref{eq:i_ez_lim} to derive constraints on $p_{e}$ as follows,
\beq\label{eq:i_ez_lim}
0 \le p^e \le \bar{p}^e:=n^e \bar{I}^e V_{e}(\bar{I}^e).
\eeq
\end{itemize}

\subsubsection{Electrochemical model of  fuel cell $p^f := f^f(h^f)$}

% $h^f_t = f^f(p^f_t) = H^f((P^f)^{-1}(p^f_t)) $

% h-I $I^{f} = H^f(h^f) = \frac{2F\cdot h^f}{S_{H_{2}}}$

% $\color{blue}I^f(h^f) \color{black} = \frac{2F\cdot \color{blue}h^f}{S_{H_{2}}}$

% \newline
% this can be simplified to:
% \newline
% $ I^{f}(h^f)  =  K^f \cdot h^f $ 
\begin{itemize}
\item{Current-Hydrogen Relationship}
\beq\label{eq:hi-fc-rel}
I^f = \frac{2F h^{f}}{n^f M_{H_{2}}},
\eeq
where $I^f (A)$ is the amount of current produced, $h^f (kg/s)$ is the amount of input hydrogen to the fuel cell, $n^f$ is the number of fuel cells in series, $F (C/mol)$ is the Faraday's constant, and $M_{H_{2}} (kg/mol)$ is the molar mass of hydrogen. In the modeling section, we express \eqref{eq:hi-fc-rel} as $I^f = k^fh^f/n^f$ in \eqref{eq:hf-if}. We therefore have,
\beq
k^f = \frac{2F}{M_{H_{2}}}.
\eeq
In the simulation section, we set $n^f = 300$, $M_{H_{2}} = 2 \cross 10^{-3}$ $kg/mol$, and $ k^f =$ 1316.5 $C/cm^2\cdot kg$. 

\item{Voltage-Current Relationship}

Similar to the PEM electrolyzer cell, the voltage of a PEM fuel cell can be defined as the result of the following expression. For a single PEM fuel cell, the following equation describes the voltage-current relationship,
\beq\label{eq:vFC_1}
V_{f}(I^f) = V^f_{Nernst} - V^f_{act} - V^f_{\Omega} - V^f_{con}.
\eeq
\newline
When there is no current through the cell ($I^f = 0$), $V_{f}(I^f)$ is simply equal to the open circuit voltage $V^f_{Nernst}$ which is given by
\begin{equation}\label{h-I relation}
\begin{aligned}
V^f_{Nernst} = 1.229 - 0.85 \cross 10^{-3}(T-298.15) + \\
4.31 \cross 10^{-5}\bigg(ln(p_{H_{2}}) + 0.5ln{(p_{O_{2}})}\bigg),
\end{aligned}
\end{equation}
where $p_{H_{2}}(bar)$ and $p_{O_{2}}(bar)$ are the partial pressures of hydrogen and oxygen, respectively. $T (K)$ is the operating temperature of the fuel cell. In the simulation section, we set $T =343$ $K$, $p_{H_{2}} = 1$ $bar$, and $p_{O_{2}} = 1$ $bar$. 
\newline
$V^f_{act}$ is the overvoltage due to the activation losses that describe the charge transfer rate at lower current densities and is given by
\beq\label{h-I relation}
V^f_{act} = -[\zeta_{1} + \zeta_{2}T + \zeta_{3}Tln(C_{O_{2}}) + \zeta_{4}Tln(I^f)],
\eeq
where $\zeta's$ represent parametric coefficients whose values are based on kinetic, thermodynamic, and electrochemical equations. Here, $I^f (A)$ is the current through the cell, $T (K)$ is the operating temperature, and $C_{O_{2}} (mol/cm^3)$ is the concentration of oxygen in the catalytic interface of the cathode, it is related to the partial pressure of oxygen as
\beq
C_{O_{2}} = \frac{p_{O_{2}}}{5.08 \cross 10^6 e^(\frac{-498}{T})}.
\eeq
In the simulation section, we set $\zeta_{1} = -0.948$, $\zeta_{2} = 0.00354$, $\zeta_{3} = 7.6 \cross 10^{-5}$, $\zeta_{4} = -1.93 \cross 10^{-4}$.
The ohmic losses caused by the electrolyte resistance and contact resistance at graphite electrodes are given by
\beq
V^f_{\Omega} = I^f (R_{C} + R_{M}),
\eeq
where $R_{C} (\Omega)$ represents the resistance to the transfer of protons and is usually considered constant.  $R_{M} (\Omega)$ is the equivalent resistance of the membrane and can be calculated as
\beq
R_{M} = \frac{\rho_{M} l }{A^f}, 
\eeq
where $\rho_{M} (\Omega \cdot cm)$ is the specific resistivity of the cell, $l (cm)$ is the thickness of the membrane and $A^f (cm^2)$ is the active surface area of the cell. In the simulation section, we set $R_{C} = 0.0003$ $\Omega$, $A^f = 232$ $cm^2$, $l = 0.0178$ $cm$, and $\rho_{M} = 9.5$ $\Omega \cdot cm$.

%\newline
Finally, the concentration overvoltage is described by
\beq\label{eq:vFC_2}
V^f_{con} = -B \cdot  ln\bigg(1 - \frac{J^f}{J^f_{max}}\bigg).
\eeq
\beq
J^f = \frac{I^f}{A^f}.
\eeq
where $J^f (A/cm^2)$ is the current density through the cell, $J^f_{max} (A/cm^2)$ is the maximum allowable current density through the cell, and $B(V)$ is the parametric coefficient that depends on the operating conditions of the cell. In the simulation section, we set $B = 0.016$ $V$ and $J^f_{max} =1.5$ $A/cm^2$.

%\newline
\item{Power-Current Relationship}:

For $n^f$ cells in series, the power-current relationship is described by
\beq\label{eq:p-i_fc}
P_{f}(I^f) = n^f I^f  V_{f}(I^f)
\eeq
where $V_{f}(I^f)$ is described by equations \eqref{eq:vFC_1}-\eqref{eq:vFC_2}. When $I^f = 0$, the cell is in an open circuit condition, implying $P_{f}(0) = 0$.
% \tcr{Cong: You need to connect $ V^f(I^f)$ with equations (23)-(29), otherwise people don't know how those equations are related to each other. I guess you need to write $J^f$ the current density equals $I^f$ divided by the area. Also you need to explain the parameter of the area you choose in your simulation. I guess you need to use $I^f=A^fJ^f$.}

\item{Operation limits}

We have the following operation limit on a single PEM fuel cell
\beq\label{eq:i_fc_lim}
0 \le I^f \le \bar{I}^{f}:=A^fJ^f_{max}.
\eeq

We can use equations \eqref{eq:p-i_fc} and \eqref{eq:i_fc_lim} to derive constraints on $p_{f}$ as follows,
\beq\label{eq:i_ez_lim}
0 \le p^f \le \bar{p}^f:=n^f \bar{I}^f V_{f}(\bar{I}^f).
\eeq

\end{itemize}

\subsection{Notations}

All the major symbols used in the main text are summarized in Table~\ref{tab:symbols}.

{\small
\begin{table}[htbp]
\caption{Major symbols}\label{tab:symbols}
\begin{center}
% \vspace{-1em}
\begin{tabular}{ll}
\hline
$T\in \mathbb{R}$& total number of 1 min time intervals\\
$N\in \mathbb{R}$& total number of households \\
$\delta$ & sampling time interval (s)\\
$v_i$& penalty for loss-of-load of customer $i$ ($\$/kW$)\\
% $W\in \mathbb{R}$& total number of 4s time intervals for AGC signals\\
$\dbf\in \mathbb{R}^{N \times T}$ & demand for a household ($kW$)\\
$\bar{d}, \underline{d} \in \mathbb{R}^T$ & upper and lower bounds on demand ($kW$)\\
$\bar{g}\in \mathbb{R}^T$ & maximum energy supply from the grid ($kW$)\\
$\overline{e}, \underline{e} \in \mathbb{R}^T$ & upper and lower bounds on hydrogen tank ($kg$)\\
$e_{0} \in \mathbb{R}$ & initial state of hydrogen in the tank ($kg$)\\
$\bar{p}^{e}, \bar{p}^{f} \in \mathbb{R}$ & maximum power of electrolyzer/fuel cell stack ($kW$)\\
% $\bar{p^f} \in \mathbb{R}$ & maximum power generated for fuel cell stack ($kW$)\\
$c_t \in \mathbb{R}$ & cost of buying energy from the grid ($\$/kW$)\\
$\eta^e, \eta^f$ & efficiency  for electrolyzer and fuel cell\\
% $\eta^f$ & constant efficiency parameter for fuel cell\\
$k^e, k^f$ & current-hydrogen factor for electrolyzer/fuel cell\\
% $k^f$ & hydrogen to current conversion factor for fuel cell\\
$n^e, n^f$ & number of cells in the electrolyzer/fuel cell stack\\
% $n^f$ & number of cells in the fuel cell stack\\
$\cal{H}$ & higher heating value of hydrogen $(kJ/kg)$\\
% $HHV$ & higher heating value of hydrogen $(MJ/kg)$\\

% $\mbf^c_i,\mbf^d_i\in \mathbb{R}^W$ & regulation up and down   mileage of unit $i$\\
\hline
\end{tabular}
\end{center}
\end{table}
}
\vspace{-0.2in}
\subsection{Mixed integer optimization for piecewise linear model}\label{sec:MILP}

We introduce notations for the breakpoints of the piecewise linear model \eqref{eq:pe_linear-he_linear}, \ie $\Pc^e_k:= [\bar{p}^e_{k}, \bar{p}^e_{k+1} ]$,  $\Hc^f_k:= [\overline{h}^f_{k}, \overline{h}^f_{k+1} ]$. At the $k$-th segment of the piecewise linear model at time $t$, $h^f_{k,t}$ and $h^e_{k,t}$ represents the hydrogen flow, and $p^e_{k,t}, p^f_{k,t}$ the power flow. We have  
\beq\label{eq:piece}
h^e_{k,t}=a^e_k p^e_{k,t} + b^e_k, ~~
p^f_{k,t}=a^f_k h^f_{k,t} + b^f_k.
\eeq

\begin{figure}[!htbp]
   \centering
   \vspace{-0.1in}
   \includegraphics[scale=0.37]{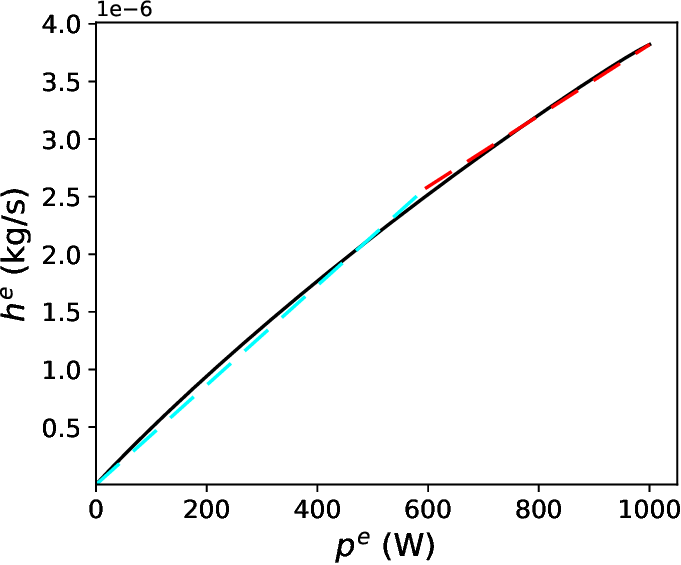} \includegraphics[scale=0.37]{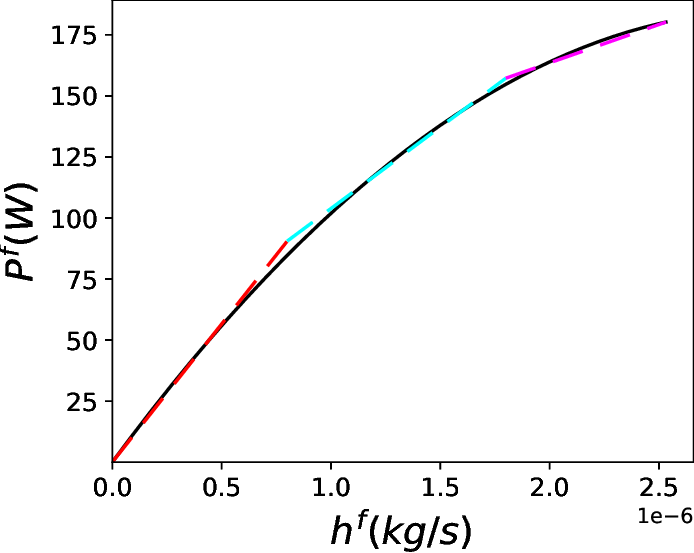}
\caption{\scriptsize  Piecewise linearized power-hydrogen relationship for a single cell. The black line is the origin nonlinear model, and the colorful dashed lines are piecewise linear approximation model. left: Electrolyzer; right: Fuel Cell.}
\label{fig:PiecesApprox}
\end{figure}
Fig.~\ref{fig:PiecesApprox} illustrates the piecewise linearization for the nonlinear power-hydrogen relationships from equations \eqref{eq:he-pe} and \eqref{eq:hf-pf}. Here, $K=2$ for the electrolyzer and $K=3$ for the fuel cell. From \eqref{eq:he-pe} and \eqref{eq:hf-pf}, it should be noted that these relationships were developed for a single PEM electrolyzer/fuel cell and their application to a stack comprising of cells connected in series can be extended by multiplying both $X$ and $Y$ axes by the number of cells $n^e$ or $n^f$. In the simulation section, we conduct the piecewise linear regression. And, for the electrolyzer, we get breakpoints $\bar{\pbf}^e = [0,595,1002]W$, $\bar{\hbf}^e =[0,2.5 \cross 10^{-6}$,$3.8 \cross 10^{-6}] kg/s$ . For the fuel cell, we get breakpoints $\bar{\hbf}^f = [0,0.8 \cross 10^{-6}, 1.8 \cross 10^{-6}, 2.5 \cross 10^{-6}] kg/s$ , $\bar{\pbf}^f = [0.90.53,157.27,180.24] W$.

We introduce binary variables $z^e_{k,t},z^f_{k,t}$ for the piecewise linear hydrogen storage model. The mixed integer optimization for microgrid with piecewise linear hydrogen storage model is given by 
\begin{subequations} 
    \begin{align}
    &  \underset{\hat{\Omega}}{\rm minimize} &&  \sum_{i=1}^N\Phi_i(\lbf_i)+\sum_{t=t'}^{T-1} c_t g_t, 
    \\
    & \text{subject to} 
    && \forall t\in \{t',..., T-1\}, i\in \{1,..., N\} , k\in \{1,..., K\}  \nn
    \\
    &&& e_{t} + (\sum_{k=1}^Kh^e_{k,t} -\sum_{k=1}^Kh^f_{k,t}) \delta =e_{t+1},
    \\
    % &&& \sum_{i=1}^N (e_{it} - l_{it})- p^f_t - g_t + p_t^e = 0,
    % \\
    &&& d_{it} - r_{it}
 - l_{it} =q_{it},\\
    &&& \sum_{k=1}^Kp^e_{k,t}  - \sum_{k=1}^Kp^f_{k,t}  - g_t = q_{0t},\\
    &&& \ul{\v{b}} \leq \v{A} \qbf_t \leq \ol{\v{b}},  \l_{it} \le d_{it},  \underline{e}  \le e_{t}\le \overline{e}, 
   \\
    &&&  g_t \le \bar{g}_t,  \underline{d}_{it} \le d_{it}\le \bar{d}_{it}, \eqref{eq:piece} \\
    &&& z^e_{k,t} \bar{p}^e_{k} \le p^e_{k,t} \le z^e_{k,t} \bar{p}^e_{k+1},  \sum_{k=1}^Kz^e_{k,t}=1, \\
    % &&& z^e_{2,t} p^e_1 \le p^e_{2,t} \le z^e_{2,t} p^e_2, \\
    &&& z^f_{k,t} \overline{h}^f_{k} \le h^f_{k,t} \le z^f_{k,t} \overline{h}^f_{k+1},  \sum_{k=1}^Kz^f_{k,t}=1,
    % &&& z^f_{2,t} h^f_2 \le h^f_{2,t} \le z^f_{2,t} h^f_2, \\
    % &&& z^f_{3,t} h^f_2 \le h^f_{3,t} \le z^f_{3,t} h^f_3,
    \end{align}
    \label{eq:EM_MIO}
\end{subequations}
where $\hat{\Omega} := \Omega\times \{z^e_{k,t},z^f_{k,t} \in {\cal B}\}$.

\subsection{Feasibility projection method}\label{sec:FP}
To correct infeasible dispatch, we adjust the hydrogen schedules based on the nonlinear model in each receding window. We propose the  feasibility projection method in Algorithm~\ref{alg:Proj} to project the infeasible solution to the feasible domain in optimization \eqref{eq:EM_multi} when conducting MPC. We denote the projected feasible solutions as  $\tilde{p}^e_t, \tilde{p}^f_t$.

\begin{algorithm}[htbp]
 \caption{MPC + Greedy feasibility Projection}\label{alg:Proj}
\SetAlgoLined
% \KwIn{Solution $p^e_t, p^f_t$ from the optimization with constant or piecewise linear model.}\\
% \textbf{Output:}optimal dispatch and pricing result.\\
% \KwOut{Feasible dispatch $\tilde{p}^e_t, \tilde{p}^f_t$ satisfying nonlinear constraints for electrolyzer and fuel cell.}\\
\textbf{Initialization:} set $t'=0$ and $\tilde{e}_0 = s$. \\
 \While{$t' \le T$}{
 Solve (\ref{eq:EM_multi}) with piecewise linear approximation.\\
 $h^e_{t'} = f^e(p^e_{t'}), h^f_{t'} = f^f(p^f_{t'})$.\\ $e_{t'+1} = e_{t'} + (h^e_{t'} - h^f_{t'}) \delta$.\\
  \textbf{If} $e_{t'+1} \le \underline{e}$: $\tilde{h}^f_{t'} = \frac{e_{t'}-\underline{e}}{ \delta} + h^e_{t'}$,  $\tilde{h}^e_{t'} = h^e_{t'} $.\\
\textbf{If} $e_{t'+1} \geq \overline{e}$: $\tilde{h}^e_{t'} = \frac{\overline{e} - e_{t'}}{ \delta } + h^f_{t'}$,  $\tilde{h}^f_{t'} = h^f_{t'} $. \\
\textbf{Otherwise}: $\tilde{h}^e_{t'} = h^e_{t'} $, $\tilde{h}^f_{t'} = h^f_{t'} $. \\
   $\tilde{e}_{t'+1} = \tilde{e}_{t'} + (\tilde{h}^e_{t'} - \tilde{h}^f_{t'}) \delta $.\\
   $\tilde{p}^e_{t'} = [f^e]^{-1}(\tilde{h}^e_{t'}),  \tilde{p}^f_{t'} = [f^f]^{-1}(\tilde{h}^f_{t'})$.\\
   Solve (\ref{eq:EM_multi}) with $p^e_{t'} = \tilde{p}^e_{t'} $, $p^f_{t'} = \tilde{p}^f_{t'} $, and implement the optimal solution at $t'$.\\
   $t'=t'+1$.}

   % compute the corrected loss-of-load$\tilde{L} = e_t -g_t + \tilde{p}^e_t- \tilde{p}^f_t$.

%   \eIf{condition}{
%   instructions1\;
%   instructions2\;
%   }{
%   instructions3\;
%   }
\end{algorithm}

% Only if $  \tilde{p}^e_{t'} - \tilde{p}^f_{t'} \le  \bar{g}_{t'} $, the above corrections always has a feasible solution in the last step.
% Here is the intuition why this algorithm can always guarantee a feasible solution. Ignoring the microgrid network congestion, in the first solve of (\ref{eq:EM_multi}) with piecewise linear approximation we have  $  \bar{g}_{t'}+ \sum_i r_{it'} - p^e_{t'} + p^f_{t'} \geq 0$ since the hydrogen storage is connected to the parent bus. Because of the monotonicity of the nonlinear electrochemical model,  the infeasibility projection always has $ 0 \le \tilde{p}^e_{t'} \le p^e_{t'}, 0 \le \tilde{p}^f_{t'} \le p^f_{t'}$. Therefore, the above correction always has  $  \bar{g}_{t'}+ \sum_i r_{it'} - \tilde{p}^e_{t'} + \tilde{p}^f_{t'}  \geq 0$ to  guarantee a feasible solution. \tcr{(Cong: we need $p^e_{t'}  p^f_{t'}  = 0$ from Proposition \ref{prop:nonSimulCD} for this part.)}

\subsection{Additional simulation results}\label{sec:sim}

 \begin{figure}[!htbp]
   \centering
   \vspace{-0.1in}
   \includegraphics[width=0.48\textwidth]{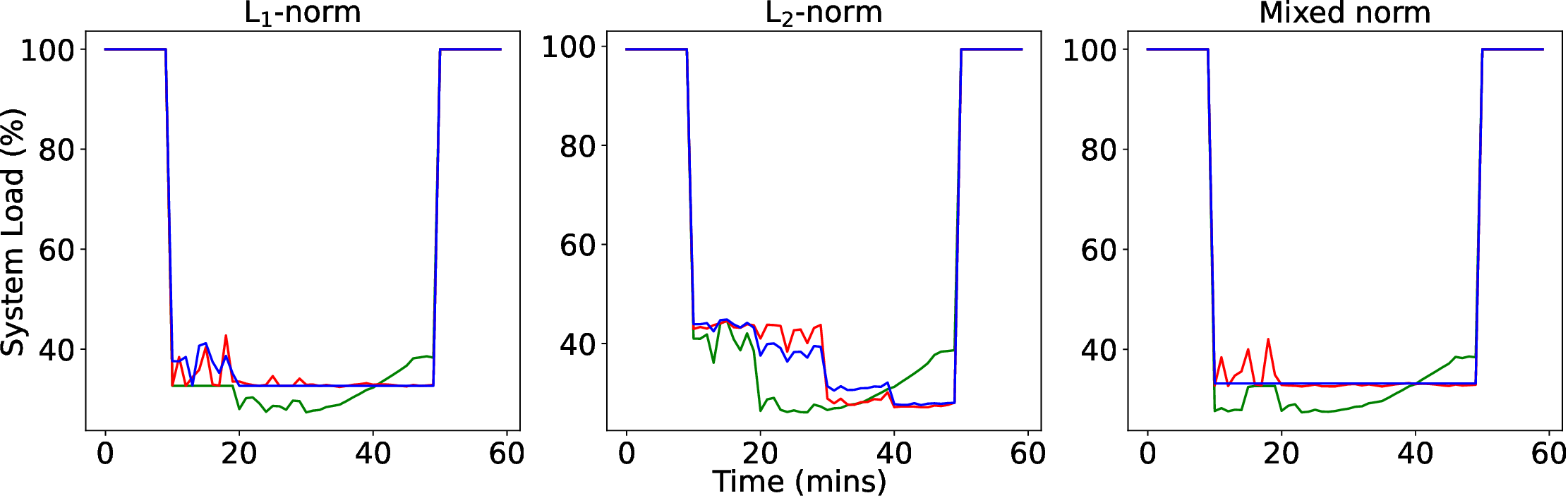}    \includegraphics[width=0.48\textwidth]{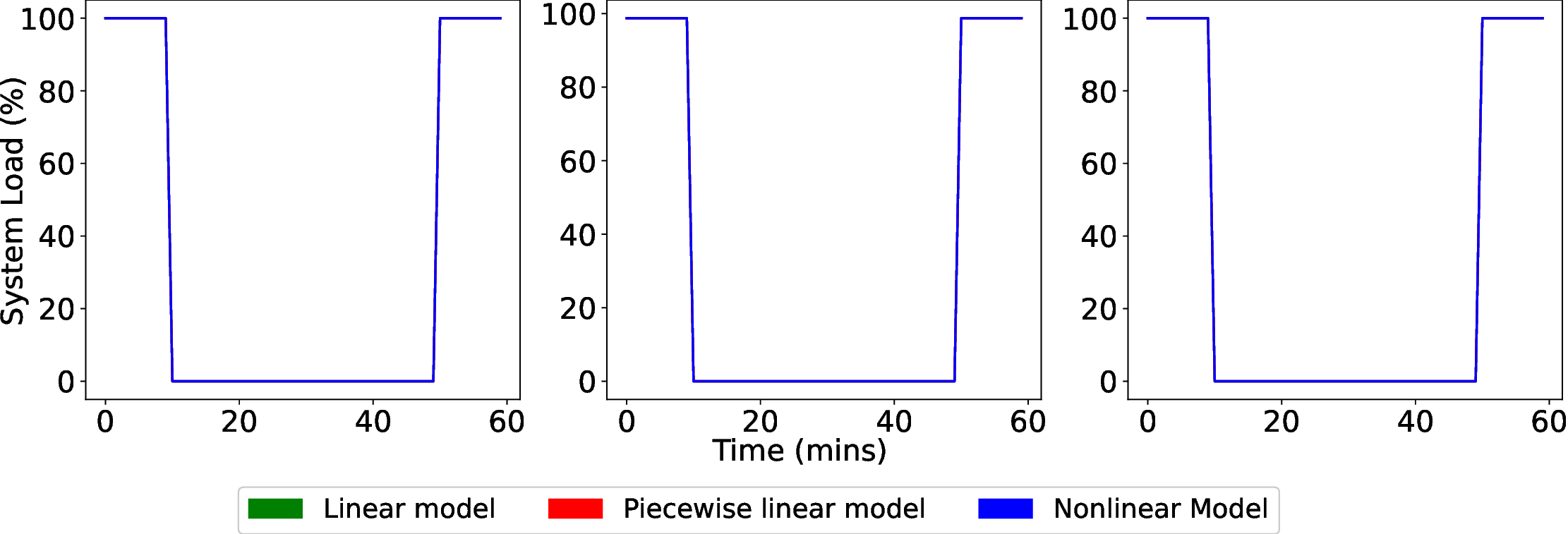}
\caption{\scriptsize  Top to Bottom:  Resilience Trapezoids for all customers and type-3 customers. Left to Right: Resilience Trapezoids under different penalty schemes.}
\label{fig:trapezoid3}
\end{figure}

In Fig.~\ref{fig:trapezoid3} (first row) we observe that the piecewise linear model reduces total loss-of-load by 3\% for all customer categories across all penalty functions compared to the linear model (green). For type-3 customers there is no difference in the performance, as we observe a complete outage for the entire duration of the contingency.

% \tcr{Cong: to control the page limit, you can put simulation results related to the core contributions in the main text, and other simulation results here.}

To further investigate the performance of different penalty functions on the duration-of-outage we solve (\ref{eq:EM_multi}) with the piecewise linear model by varying grid supply and renewable trajectories as:
$\bar{g_{t}} = \alpha \bar{g_{t}}$, $r_{it} = \alpha r_{it}$ for $\alpha \in \{0,0.2,0.4,0.6,0.8,1,2\}$. 

 \begin{figure}[!htbp]
   \centering
   \vspace{-0.1in}
   \includegraphics[scale=0.35]{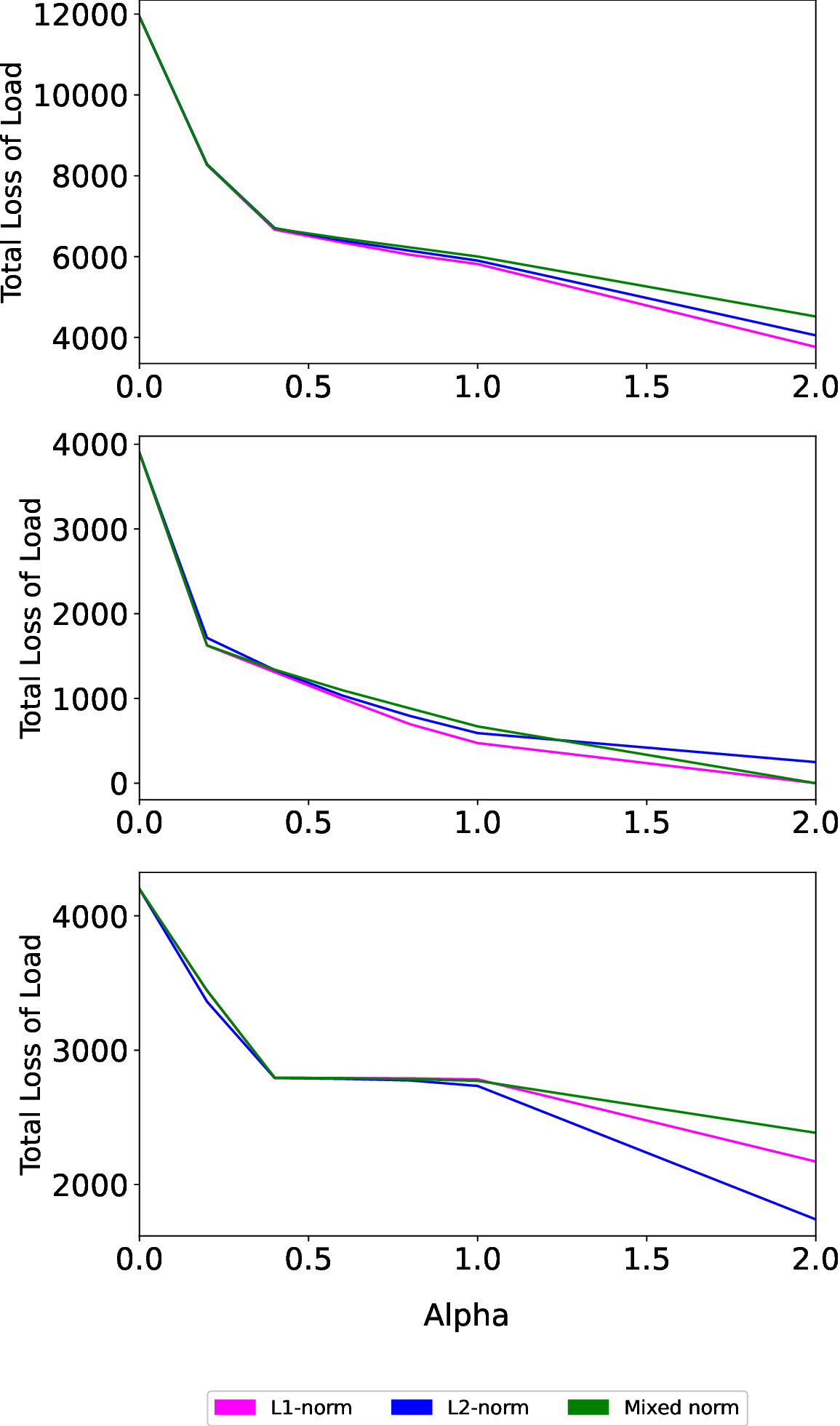}
\caption{\scriptsize  Top to Bottom: Total loss-of-load for all customers, type-1 customers, and type-2 customers. }
\label{fig:alpha_total}
\end{figure}
Fig.~\ref{fig:alpha_total} shows how the total loss-of-load under different penalty functions changes with $\alpha$. We observe that $l_{2}$-norm performs within 0\%-7.6\% range of $l_{1}$-norm in total loss-of-load as can be seen in Fig.~\ref{fig:alpha_total} (first row). It is interesting to note, the distinct manner in which loss-of-load is allocated to different customer categories under $l_{2}$-norm compared to $l_{1}$-norm and mixed norm. Fig.~\ref{fig:alpha_total} (second row) shows that loss-of-load for type-1 customers under the $l_{2}$-norm penalty significantly exceeds (by (0\%-24\%) that of $l_{1}$-norm and mixed norm. Conversely, as shown in Fig.~\ref{fig:alpha_total} (last row) the loss-of-load for type-2 customers under the $l_{2}$-norm penalty is reduced (by 0\%-27\%) in comparison to $l_{1}$-norm and mixed norm. This highlights the $l_{2}$-norm's role in promoting a more equitable distribution of loss-of-load among different customer categories.  

 \begin{figure}[!htbp]
   \centering
   \vspace{-0.1in}
   \includegraphics[scale=0.35]{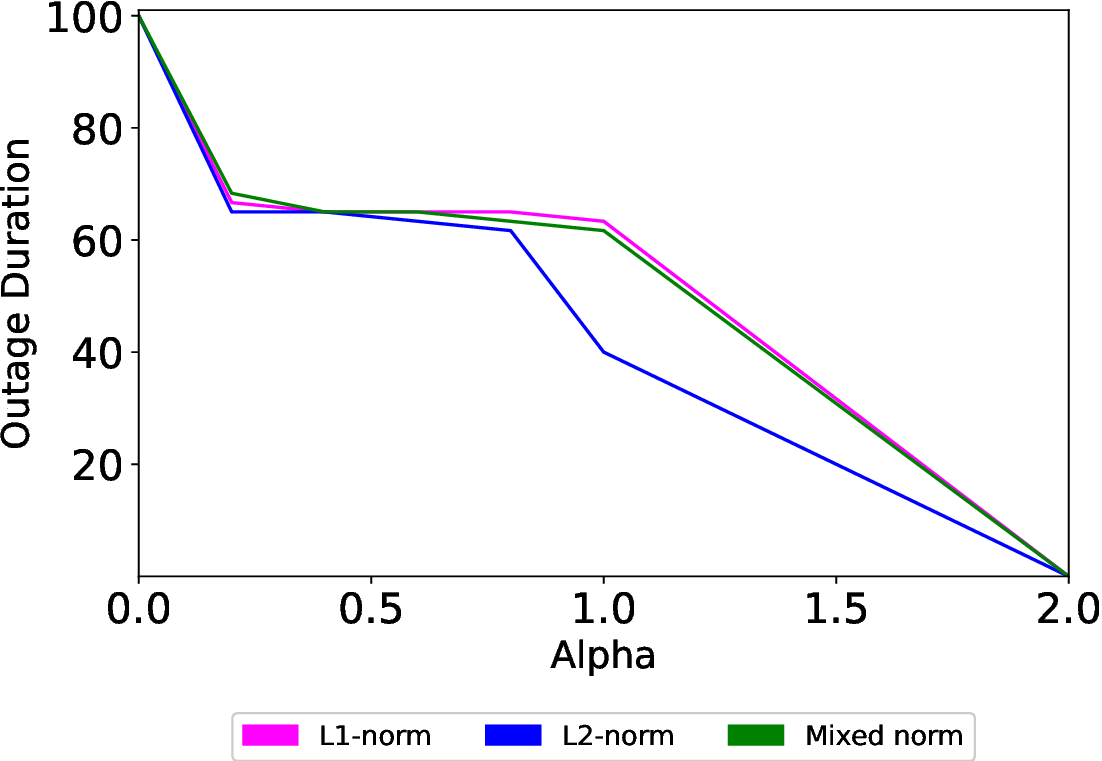}
\caption{\scriptsize Duration of outage for type-2 customers. }
\label{fig:alpha}
\end{figure}

Fig.~\ref{fig:alpha} shows how the outage duration, as a \% of the total time horizon changes with $\alpha$.
We observe that for type-2 customers, $l_{2}$-norm consistently outperforms $l_{1}$-norm and mixed norm penalty functions. We observe a maximum reduction of 37\% for $\alpha = 1$. This can be attributed to how the total lost load is distributed amongst different categories of customers, illustrated in Fig.~\ref{fig:alpha_total}. A lesser loss-of-load leads to a reduction in the duration-of-outage for type-2 customers.

\end{document}